\documentclass[11pt]{article}

\usepackage[T1]{fontenc}
\usepackage[utf8]{inputenc}
\usepackage{lmodern}
\usepackage{textcomp}
\usepackage{amsmath,amssymb,amsfonts}
\usepackage{bm}
\usepackage[a4paper,margin=1in]{geometry}
\usepackage{xcolor}
\usepackage{graphicx}
\usepackage{longtable,booktabs,array}
\usepackage{calc}
\usepackage{etoolbox}
\usepackage{microtype}
\usepackage[normalem]{ulem}
\usepackage{parskip}
\usepackage{newunicodechar}
\usepackage[colorlinks=true,allcolors=blue!55!black,breaklinks=true]{hyperref}
\usepackage{xurl}

\makeatletter
\patchcmd\longtable{\par}{\if@noskipsec\mbox{}\fi\par}{}{}
\makeatother

\makeatletter
\def\maxwidth{\ifdim\Gin@nat@width>\linewidth\linewidth\else\Gin@nat@width\fi}
\def\maxheight{\ifdim\Gin@nat@height>\textheight\textheight\else\Gin@nat@height\fi}
\makeatother
\setkeys{Gin}{width=\maxwidth,height=\maxheight,keepaspectratio}

\providecommand{\ul}[1]{\uline{#1}}
\newenvironment{refslist}%
  {\par\small\sloppy\setlength{\parindent}{0pt}\setlength{\parskip}{4pt plus 1pt}%
   \everypar{\hangindent=1.5em\hangafter=1}}%
  {\par}

\newunicodechar{ }{\,}
\newunicodechar{‐}{-}
\newunicodechar{–}{--}
\newunicodechar{“}{``}
\newunicodechar{”}{''}
\newunicodechar{•}{\textbullet{}}
\newunicodechar{ˉ}{\ifmmode\bar{}\else\textasciimacron\fi}
\newunicodechar{°}{\textdegree{}}
\newunicodechar{±}{\ensuremath{\pm}}
\newunicodechar{×}{\ensuremath{\times}}
\newunicodechar{·}{\ensuremath{\cdot}}
\newunicodechar{⋅}{\ensuremath{\cdot}}
\newunicodechar{½}{\textonehalf{}}
\newunicodechar{√}{\ensuremath{\surd}}
\newunicodechar{∝}{\ensuremath{\propto}}
\newunicodechar{∼}{\ensuremath{\sim}}
\newunicodechar{≈}{\ensuremath{\approx}}
\newunicodechar{≠}{\ensuremath{\neq}}
\newunicodechar{≥}{\ensuremath{\geq}}
\newunicodechar{−}{\ensuremath{-}}
\newunicodechar{∂}{\ensuremath{\partial}}
\newunicodechar{∇}{\ensuremath{\nabla}}
\newunicodechar{′}{\ensuremath{{}^{\prime}}}
\newunicodechar{↑}{\ensuremath{\uparrow}}
\newunicodechar{↓}{\ensuremath{\downarrow}}
\newunicodechar{→}{\ensuremath{\rightarrow}}
\newunicodechar{↔}{\ensuremath{\leftrightarrow}}
\newunicodechar{¹}{\ensuremath{^{1}}}
\newunicodechar{²}{\ensuremath{^{2}}}
\newunicodechar{³}{\ensuremath{^{3}}}
\newunicodechar{⁴}{\ensuremath{^{4}}}
\newunicodechar{⁵}{\ensuremath{^{5}}}
\newunicodechar{⁶}{\ensuremath{^{6}}}
\newunicodechar{⁷}{\ensuremath{^{7}}}
\newunicodechar{⁹}{\ensuremath{^{9}}}
\newunicodechar{⁺}{\ensuremath{^{+}}}
\newunicodechar{⁻}{\ensuremath{^{-}}}
\newunicodechar{₀}{\ensuremath{_{0}}}
\newunicodechar{₁}{\ensuremath{_{1}}}
\newunicodechar{₂}{\ensuremath{_{2}}}
\newunicodechar{Γ}{\ensuremath{\Gamma}}
\newunicodechar{Δ}{\ensuremath{\Delta}}
\newunicodechar{Θ}{\ensuremath{\Theta}}
\newunicodechar{α}{\ensuremath{\alpha}}
\newunicodechar{β}{\ensuremath{\beta}}
\newunicodechar{γ}{\ensuremath{\gamma}}
\newunicodechar{δ}{\ensuremath{\delta}}
\newunicodechar{η}{\ensuremath{\eta}}
\newunicodechar{κ}{\ensuremath{\kappa}}
\newunicodechar{λ}{\ensuremath{\lambda}}
\newunicodechar{μ}{\ensuremath{\mu}}
\newunicodechar{µ}{\ensuremath{\mu}}
\newunicodechar{σ}{\ensuremath{\sigma}}
\newunicodechar{τ}{\ensuremath{\tau}}
\newunicodechar{φ}{\ensuremath{\phi}}
\newunicodechar{ϕ}{\ensuremath{\phi}}
\newunicodechar{ψ}{\ensuremath{\psi}}
\newunicodechar{á}{\'a}
\newunicodechar{ä}{\"a}
\newunicodechar{è}{\`e}
\newunicodechar{é}{\'e}
\newunicodechar{ö}{\"o}
\newunicodechar{ü}{\"u}
\newunicodechar{ß}{\ss{}}
\newunicodechar{Ă}{\u{A}}
\newunicodechar{ă}{\u{a}}

\begin{document}

\begin{center}
{\LARGE\bfseries Neuronal electricality founded in murburn-thermodynamic principles}\\[5pt]
{\large\bfseries Background and basic theoretical formulation}\\[6pt]
{\large \emph{Kelath Murali Manoj\textsuperscript{1,2}, N Sukumar\textsuperscript{1}, Taufia Hussain}\textsuperscript{3}\emph{, Mahendra Kavdia}\textsuperscript{4}\emph{, Abhijith Anandakrishnan}\textsuperscript{1}}\\[8pt]
{\footnotesize \textsuperscript{1} Amrita School of Artificial Intelligence, Coimbatore, Amrita Vishwa Vidyapeetham, Amritanagar, Ettimadai 641112, Tamil Nadu, India. \\[2pt]
\textsuperscript{2} Satyamjayatu: The Science \& Ethics Foundation, Shoranur-2, Palakkad Dist. 679122, Kerala, India. \\[2pt]
\textsuperscript{3}Department of Animal Physiology, Institute of Biosciences, Albert-Einstein-Straße 3, University of Rostock, 18059 Rostock, Germany \\[2pt]
\textsuperscript{4}Department of Biomedical Engineering, Wayne State University, \\[2pt]
Detroit, Michigan, USA \\[2pt]
*KMM: km\_manoj@cb.amrita.edu; murman@satyamjayatu.com; ORCID: 0000-0003-4515-994X \\[2pt]
NS: n\_sukumar@cb.amrita.edu; ORCID: 0000-0002-2724-9944 \\[2pt]
TH: taufia.hussain@uni-rostock.de; ORCID: 0009-0000-8334-4762 \\[2pt]
MK: kavdia@wayne.edu \\[2pt]
AA: a\_abhijith@cb.amrita.edu; ORCID: 0000-0002-4650-8438}
\end{center}

\begin{abstract}
\noindent Trans-membrane ion-gradients and fluxes (H\textsuperscript{+}, Na\textsuperscript{+}, K\textsuperscript{+}, etc.) are central to conventional electrical activity in aerobic cells/organelles. Murburn concept (an umbrella of theorization based in stochastic redox processes) offers novel physico-chemical perspectives/models for various metabolic, bioenergetic and electrophysiological phenomena. Here, we develop a foundational framework for neuronal electrical activity and axonal signal propagation using the electron-holding potential (EHP, ϕ), a dimensionless field related logarithmically to electron chemical potential. By combining local redox relaxation dynamics with spatial transport driven by thermodynamic gradients, we derive a unified reaction-transport-relaxation equation that accounts for resting potential, excitability, waveform generation, and signal propagation within a single formalism. Nonlinear local redox kinetics yield a stable resting state and graded responses from a single scalar field; extending it to the two-variable excitable (FitzHugh--Nagumo) form, a bistable reaction with a slow recovery variable, further yields a genuine threshold, all-or-none spikes, a refractory period and a propagating action potential. The framework accommodates known physiological variability of neurons and provides a direct bridge between metabolic/redox state and electrophysiology. This framework offers testable predictions for neuronal dynamics (with respect to observed or alterable features like: velocity, waveform morphology, environmental/physiological conditions, etc.) across biological systems. Here we derive and solve the equations to obtain the transmembrane potential as function of time, and the neuronal conduction velocity as a function of various parameters such as ionic strength, temperature axon diameter and myelination, and driving potential. In the second part of this work, we present comparative analyses, simulations, and experimental strategies for validation and falsification.

\medskip\noindent\textbf{Keywords:} murburn concept, classical membrane pump theory, Goldman-Hodgkin-Katz relation, Hodgkin-Huxley model, neuronal electricality
\end{abstract}

\section*{1. Background to the context of electrophysiology}

The classical membrane pump theory (CMPT) describes neuronal signalling as a consequence of selective ionic permeability and voltage-gated trans-membrane cation-fluxes (Hodgkin \& Huxley, 1952; Neher \& Sakmann, 1976; Kandel et al., 2021). We argue that while such formulations elegantly reproduce observed voltage waveforms phenomenologically and agree with many experimental observations qualitatively, while several deeper mechanistic questions remain open. Sometimes, a mechanistic rationale that is consistent with fundamental chemical kinetics, thermodynamics, electrostatics and with the known structural features of proteins/membranes are not easily available (Manoj et al., 2026c). Further, several physiological observations were incompatible/inexplicable with the classical perceptions. For example, while the theory demands that the in-out concentrations of the major cations lead to resting trans-membrane potential (TMP), the mechanistic explanation relies on spatio-temporally and contextually changing permeability constants to account for the relatively unpredictable TMP values of diverse systems. (That is, the observed concentration gradients of the ions cannot explain TMP on their own merit!) Also, key theoretical aspects, particularly the energetic cost, speed, and robustness of neuronal impulse conduction remain difficult to reconcile with pump \& channel--centric ionic-models. As a consequence of these lacunae, and in order to accord with various aspects of neuronal impulse conduction, several models such as ``oscillators'', ``solitons'' and ``electromagnetic waves'' were proposed by many researchers over the years (FitzHugh, 1961; Heimburg \& Jackson, 2005; El Hady \& Machta, 2015).

Our pursuits had pointed out that neither the older nor the later models could afford a general connectivity/continuum with basic cellular biophysical chemistry, and we extended the murburn-umbrella of cellular redox metabolism (Manoj et al., 2016; Gideon et al., 2022; Manoj et al., 2022a) and bioenergetics (Manoj \& Bazhin, 2021; Parashar et al., 2022) to explain electrophysiology also (Manoj \& Tamagawa, 2022; Manoj et al., 2022b-c; Manoj et al., 2023a-c; Manoj \& Jaeken, 2023; Jaeken \& Manoj, 2025). This new perspective was also recently grounded in mathematical and algorithmic (Manoj et al., 2026a-c) logic for catalytic and electrical-activity contexts. In the latest of these articles (Manoj et al., 2026c), we had revisited the GHK relation and Hodgkin-Huxley model (and interpretations thereof) and demonstrated that there were many mechanistically questionable aspects pertaining to the physiology of neurons. In particular, the structure-function correlations of proteins like NKA (or Na/K-ATPase, the first of the so-called electrogenic ion-pumps) and the potassium channel of KcsA (believed to be a selective and high throughput membrane channel) agreed more with an interfacial murzyme role for active/passive redox sifting of cations. Further, a simple quantitative relation (based on anionic diffusible reactive species or ADRS) for observed TMP was presented therein.

In continuation, here, we present a novel/alternative mathematical theory for neuronal electrical activity (electricality) based on murburn concept. In this perspective, ECS (effective charge separation), DRS (diffusible reactive species), and CEM (chemico-electromagnetic matrix), are the new principles/protagonists that enable cellular PCHEMS (powering-coherence-homeostasis-electromechanical \& sensing-response activities), fashioning the cells as SCE (simple chemical engines). Here, we provide even more arguments against the ion-centric classical perceptions of neuronal signalling. Further, TMP fluctuations during neuronal activity and neuronal conduction velocity (NCV) are reformulated as a directional redox process governed by Electron Holding Potential (EHP, to be elaborated later) of cellular components. In this view, one-electron chemistry in a screened ionic medium provides the causal drive, redox proteins like neuroglobin serve as one-electron buffers/pit-stops, and ions primarily serve for electrostatic screening and osmotic buffering roles; whereas some minimal fluxes of cations across axolemma may be associated with the movement of electrons along the same.

\noindent\textbf{Nomenclature.} Principal abbreviations used in this part:

\smallskip
{\small
\begin{longtable}[]{@{}ll@{}}
\toprule
Acronym & Expansion \\
\midrule
\endhead
\bottomrule
\endlastfoot
CMPT & Classical membrane pump theory \\
GHK & Goldman--Hodgkin--Katz (relation) \\
HH & Hodgkin--Huxley (model) \\
TMP & Trans-membrane potential \\
NCV & Neuronal conduction velocity \\
NKA & Na/K-ATPase \\
ECS & Effective charge separation \\
DRS & Diffusible reactive species \\
ADRS & Anionic diffusible reactive species \\
CEM & Chemico-electromagnetic matrix \\
EHP ($\phi$) & Electron-holding potential (the state variable) \\
EDP & Electron-donating property (liberating / generating) \\
ERP & Electron-relaying property (conducting / amplifying) \\
ESP & Electron-sinking property (terminating / stabilizing) \\
PCHEMS & Powering, coherence, homeostasis, electro-mechanical and sensing-response activities \\
SCE & Simple chemical engine \\
\end{longtable}
}

\section*{2. Revisiting the CMPT briefly to establish the context for a new model}

The mathematical heart of CMPT\textquotesingle s explanation for the resting TMP is the Goldman-Hodgkin-Katz (GHK) equation, which correlates the TMP with the concentration gradients and relative permeabilities of the major bulk ions (Na⁺, K⁺, Cl⁻). In doing so, it treats both bulk electrolytic behavior and surface electrostatic capacitive phenomena. The major assumptions (A1 to A7) are:

A1. Constant-field assumption: Electric field across membrane is spatially uniform.

A2. Continuum electro-diffusion: Ion transport described by Nernst--Planck flux.

A3. Independent ion movement: Each ionic species moves independently.

A4. Bulk electroneutral reservoirs: Inside and outside compartments are electroneutral.

A5. Passive steady state: No explicit chemical reactions generating charge.

A6. The `phenomenological' permeability: Ion-flux regulated by coefficients of selectivity.

A7. Voltage as causal (HH): Membrane voltage treated as the primary variable governing flux.

Now, the basic features of the GHK relation are:

Bulk electrolyte behavior (the inputs): The equation uses the concentrations of ions like Na⁺ and K⁺, which are present at high quantities (\textasciitilde10\textsuperscript{−1} M or 100 mM). These are bulk phase properties. It treats the inside and outside of the cell as two large, well-stirred reservoirs of electrolyte solution. The thermodynamic force for diffusion is the chemical potential gradient derived from these bulk concentrations.

Surface electrostatic/capacitive behavior (the mechanism \& output): The equation aims to predict a voltage (the TMP), which is a surface potential. This voltage is established across a capacitor (the lipid bilayer membrane), which is only \textasciitilde5 nm thick and has a very high capacitance. The key insight from electrostatics is that the voltage across a capacitor is determined by a tiny separation of charge, not by bulk concentrations.

A quick calculation shall illustrate the bulk and surface-ionic features:

Consider a spherical cell of radius 10 µm. Its surface area is \textasciitilde1.256×10\textsuperscript{-9} m².

\begin{quote}
The capacitance of a biological membrane is \textasciitilde0.01 F/m². So, the cell\textquotesingle s total capacitance (C) is \textasciitilde1.26×10\textsuperscript{-11} F.

The observed TMP voltage (V) is -0.07 V. The charge (Q) separated to create this voltage is given by Q = C * V.
\end{quote}

Q = (1.256×10\textsuperscript{-11} F) × 0.07 V = 8.8×10\textsuperscript{-13} Coulombs.

The charge of a single K⁺ ion is 1.6×10\textsuperscript{-19} C.

\begin{quote}
Therefore, the number of excess K⁺ ions needed outside to create this -70 mV potential is: N = 8.8×10\textsuperscript{-13} / 1.6×10\textsuperscript{-19} ≈ 5,500,000 ions.
\end{quote}

This sounds like a lot, but compare it to the bulk totals:

Cell volume is \textasciitilde4.2×10\textsuperscript{-15} m³, or 4.2×10\textsuperscript{-12} liters.

\begin{quote}
At a bulk K⁺ concentration of 140 mM (0.14 moles/liter), the total number of K⁺ ions inside the cell is: (0.14 mol/L) × (6.022×10\textsuperscript{23} ions/mol) × (4.2×10\textsuperscript{-12} L) ≈ 3.5×10\textsuperscript{11} ions.
\end{quote}

In GHK, the TMP is created by an imbalance of only about 0.0016\% of the total K⁺ ions in the cell. The bulk concentration of K⁺ would barely register a change if this imbalance were corrected. The GHK relation predicts voltage as a permeability-weighted function of ionic concentration gradients, such that membrane potential depends critically on relative trans-phase ionic flux. That is, the proposal of selective ionic permeabilities (or affinity-enabled fluxes, a deterministic feature of the membrane proteins) are its mainstay. The GHK-HH formalism addresses how voltage behaves given ionic gradients and conductances, but does not consider the deeper chemical processes that generate and maintain the charge asymmetry underlying those gradients. That the required imbalance is tiny is, however, exactly what the capacitor picture itself predicts, so the small surface-charge fraction is not on its own a defect of the classical treatment. Our reservation is interpretive rather than quantitative: the GHK-HH description is phenomenological about the chemical processes that generate and maintain the charge asymmetry, even where its charge budget is internally consistent.

Let's see the CMPT from another angle. In the Hodgkin-Huxley (HH) framework, the action potential upstroke is caused by Na⁺ influx and the repolarization is caused by K⁺ efflux. Therefore, the charge needed to change the membrane voltage must be supplied by transmembrane ionic currents, and with this, the model implicitly assumes that enough Na⁺ and K⁺ ions cross the membrane during a single spike to account for the observed change in membrane charge. Let's see how much charge is actually required to change V\textsubscript{TMP}, and to start off, let's assume that membrane behaves approximately as a capacitor.

\begin{quote}
Typical neuronal membrane capacitance, \emph{C\textsubscript{m}} ≈ 1 μF cm\textsuperscript{−2}.

As \emph{Q = CΔV} and Farad x Volt = Coulomb, for a voltage change of \textasciitilde100 mV (as the resting voltage changes from -70 mV to +30 mV), Q = (1×10\textsuperscript{-6} F cm\textsuperscript{-2}) (0.1 V) = 1×10\textsuperscript{-7} C cm\textsuperscript{-2}

As 1 Coulomb = 6.24 x 10\textsuperscript{18} electrons, the above approximates to 6.2 x 10\textsuperscript{11} ions per cm\textsuperscript{2}. (With Avogadro's number, this amounts to \textasciitilde10\textsuperscript{-12} moles (or 1 pmole) per cm\textsuperscript{2}.

So, HH implicitly requires only \textasciitilde10\textsuperscript{12} charges' cross-membrane movement per cm² per spike.
\end{quote}

The measured movement is about 0.82 to 1.54 pmol Na⁺/cm² (Carter \& Bean, 2009). This, although not inadequate, is only a minute fraction of the total ionic content inside (\textgreater0.1 moles/dm\textsuperscript{3}). The issue raised here is not one of arithmetic sufficiency, but of deeper physicochemical origin/relevance. Indeed, this measured entry of \mbox{\(\sim\)1} pmol Na\textsuperscript{+}/cm\textsuperscript{2} matches the capacitive charge that HH requires, so the agreement supports the HH charge budget rather than undermining it. What stays open is interpretive: HH builds on capacitive current mathematically but assigns the causal drive to bulk electrolytic ions, whereas the question we raise is whether that ionic flux is the cause or a downstream correlate of redox-mediated charge redistribution.

In the Hodgkin--Huxley (HH) framework, action potential propagation is described as a consequence of transmembrane ionic influx through voltage-gated channels and axial ionic current flow along a conductive cytoplasm, as formalized later in classical cable theory. The cable equation models the axon as a distributed RC circuit consisting of a resistive intracellular core and a capacitive membrane. In this representation, depolarization at one segment produces local electric fields that drive ionic drift currents, which in turn depolarize adjacent segments.

It is useful to distinguish clearly between diffusion and drift. The root-mean-square displacement for a freely diffusing inorganic ion in one dimension over time \emph{t} is given by the Einstein--Smoluchowski relation {[}\emph{Δx} = (2\emph{Dt})\textsuperscript{½}{]}. For diffusion coefficients on the order of 10\textsuperscript{−9} m\textsuperscript{2}/s, the characteristic displacement over one second is \textless100 µm. However, signal propagation in axons occurs at velocities ranging from \textasciitilde0.2 to 20 m/s, and in large myelinated fibers up to \textasciitilde200 m/s (DeMaegd et al., 2017). As in the case of acoustic waves (where wave velocity greatly exceeds individual molecular velocities) the propagation speed of a signal need not equal the microscopic particle displacement speed. Thus, classical electrophysiology relies on local field-driven drift and capacitive charge redistribution. In this ``local circuitry'' description, sodium influx at one site produces a small local change in charge density. This creates an axial electric field that drives ionic drift currents over nanometer-scale displacements, sufficient to perturb nearby voltage-gated channels. Importantly, this mechanism does not create new charge; it redistributes existing ionic charge within a highly conductive medium. From this standpoint, the axon behaves as a lossy RC (resistance-capacitance) transmission line rather than as a conduit of long-range ionic diffusion. The centrality of Na⁺ influx as the deterministic and universal initiator of action potential upstroke is strongly supported in classical axonal systems, including sodium substitution experiments (Fatt \& Katz, 1953). Importantly, Na-influx (electro-diffusion leading to an ionic-drift thereof) is seen as the fundamental depolarizing drive for signal conduction in CMPT.

Nevertheless, several conceptual questions arise when one considers the full physiological context. The classical framework treats membrane voltage both as a result of charge redistribution and as the causal variable that drives channel gating. This raises a deeper physical question: what constitutes the primary origin of the initial perturbation that destabilizes the resting state? In classical perspective, the energy source is attributed to electrochemical gradients maintained by ATP-dependent pumps and in turn, the ATP-synthesis is also resulting out of a chemico-electrical gradient (circular logic). We have proposed alternative murburn foundations for redox metabolism (Manoj et al., 2016), ATP-synthesis (Manoj \& Bazhin, 2021) and ion-gradient establishment (Manoj et al., 2022b--c, 2023b--c), suggesting that these processes are well-governed by thermodynamic mandates hitherto overlooked. The murburn rationales in these contexts excel the classical perceptions at all explanatory fronts (Jaeken \& Manoj, 2025; Manoj et al., 2026a-c). For the first time, a non-circular `cause-consequence' logic and solid foundation for seamless cellular PCHEMS (powering, coherence, homeostasis, electro-mechanical and sensing-response) activities is availed. Also, alternative ionic mechanisms have been demonstrated in other excitable membranes. Intracellular K⁺ channel blockade (Armstrong, 1971; Cahalan, 1978) markedly alters spike waveform and amplitude, indicating that depolarization dynamics do not arise from a single ionic species in isolation. Moreover, Ca²⁺-mediated action potentials in certain neuronal and cardiac preparations (Llinás \& Sugimori, 1980; Mangoni \& Nargeot, 2008; Vandael et al., 2013) demonstrate that sodium is not universally required for excitability across biological systems. Also, the refractory period is modelled in HH theory through time-dependent channel inactivation kinetics. While this provides a phenomenological description consistent with voltage-clamp data, it does not explicitly incorporate metabolic, redox, or oxygen-dependent constraints that may influence recovery dynamics in living systems. In a recent article (Manoj et al., 2026c), we had discussed the thermodynamic/kinetic basis and direct evidence for anionic diffusible reactive species (ADRS) gradients (and its relation with TMP) and also discussed the compatibility of mainstream perceptions with patch-clamp data and channel genetics, further exposing the unexplained aspects in the field.

In the early electrophysiological experiments, including Na⁺/K⁺ tracer studies, the ionic fluxes were correlated with voltage changes. However, correlation does not uniquely establish causation; it remains conceivable that measured ionic movements reflect downstream responses to more fundamental redox-mediated charge redistribution processes occurring within the cellular milieu. From a physicochemical perspective, capacitive charge rearrangement can arise not only from ionic movement but also from electronic polarization, redox-mediated charge displacement, and reorientation of interfacial dipoles in structured aqueous environments. The HH model describes voltage changes in terms of ionic fluxes and membrane capacitance but does not explicitly incorporate oxygen sensitivity, ionic-strength dependence beyond conductivity, or bulk-phase redox charge-generation mechanisms. Classical spike dynamics depend sensitively on interplay between inward and outward currents. This interplay is also modulated by redox or metabolic factors, and the GHK-HH (CMPT) formalism does not explicitly incorporate these couplings.

Thus, it is quite reasonable to question whether HH-Cable theory fully captures the chemico-electromagnetic complexity of living neurons and this is why multiple alternate explanations for neuronal impulse conduction exist in the field. If cellular electrical activity is coupled to continuous redox processes, metabolic flux, and diffusible reactive species (DRS), then neuronal electrophysiology cannot be considered an isolated ionic phenomenon. Recent developments in cellular redox biology (particularly, murburn concept) motivate a re-examination of neuronal electrical behavior within a broader chemical--electronic framework.

\section*{3. Murburn model for the resting and active electricality of neurons}

Murburn concept differs from the GHK assumptions by espousing (and questioning): chemical charge generation (A5), non-uniform fields (A1), derived (not deterministic or causal) permeability (A6), correlated ion behavior (A3), continuum limitation (A2), and voltage causality reversal (A7). It deems that effective charge separation (ECS) and diffusible reactive species (DRS) can lead to generation of transient charge asymmetries and heterogeneous fields, thereby rendering membrane voltage an emergent consequence of chemico-electronic dynamics (rather than serving as the primary driver of ionic transport). The most fundamental murburn postulate is that primarily, cellular electrical activity arises from the generation of anionic diffusible reactive species (or ADRS) within. In aerobes, it is manifested by the oxygen-centred production of superoxide, hydroxide and other peroxide-derived (soluble-hydrated) or membrane-harboured species (like partially reduced ubiquinones or CoQ\textsuperscript{*-}) or other transiently stabilized (in)organic anions. The murburn perception is fundamentally electronic and pan-systemic; that is- oxygen's electronegativity carries out effective charge separation (or ECS, generating DRS), and sets in a chemico-electromagnetic matrix (CEM). The phospholipid membrane also exercises (within the dielectric constraints) repulsion to negatively charged anions, retaining them inside and the respiratory ions inside would prefer to have a higher amount of potassium ions (as compared to sodium), owing to thermodynamic dictates (Manoj et al., 2022b, 2023a). This gives the observed resting TMP and ionic differentials, and within this purview, the major cations are not the primary driving agents for the observed electrical activity (although they play secondary roles in electrical resetting of cellular functions). That is, unlike the need for an endergonic and electrogenic cation-pumping activity in the earlier school (resulting in a lowering of positive charge inside; thereby leading to a negative TMP), the murburn model posits the more direct and simpler formation/stabilization of extra negative charges inside (resulting in a relatively negative TMP), and accommodates both bulk liquid and surface distribution of ADRS. This is largely possible with the presence of molecules like NADH inside, which generates a positive charge only with the release of two electrons. A unified and holistic derivation (a single equation meeting all aspects of neuronal electricality) of the overall electricality of neurons follows.

\subsection*{I. Definition of the state variable}

We introduce a field variable, EHP (electron holding potential), as the effective tendency of a species or microenvironment to retain electron(s)/electrical charges:

\[\boxed{\phi(x,t) \equiv \frac{a(x,t)}{a_{\text{ref}}}}\]

\(a(x,t)\) = activity (not mere concentration) of all electron-bearing species \(i\) at position \(x\), time \(t\), defined dimensionlessly as \(a(x,t) = \sum_{i}\gamma_{i}(x,t)\,c_{i}(x,t)/c^{\circ}\); here \(c_{i}\): concentration, \(\gamma_{i}\): activity coefficient (dimensionless), and \(c^{\circ}\): standard reference concentration. Making \(a\) (hence \(\phi = a/a_{\text{ref}}\)) dimensionless from the outset keeps the argument of every logarithm dimensionless.

\(a_{\text{ref}}\ \)= constant reference activity (same unit as \(a)\) ; and

\(\phi(x,t)\) is dimensionless electron holding field at position \(x,\) time \(t\).

The field variable \(\phi\) is grounded in the thermodynamic relation by considering the chemical potential of electrons at position \(x,\) time \(t\):

\[\boxed{\mu_{e}(x,t) = \mu_{e}^{0} + RT\ln a(x,t)}
\]

where- \(\mu_{e}(x,t)\) is electron chemical potential (J·mol\textsuperscript{-1}); \(\mu_{e}^{\circ}\): standard electron chemical potential (reference state); \(R\): universal gas constant (8.314 J·mol\textsuperscript{-1}·K\textsuperscript{-1}); \(T\): absolute temperature (K).

Since:

\[a(x,t) = a_{ref} \cdot \phi(x,t)
\]we get:

\[\mu_{e}(x,t) = \mu_{e}^{\circ} + RT\ln a_{ref} + RT\ln\phi(x,t)
\]Defining a new constant:

\[\boxed{\mu_{e}^{ref} \equiv \mu_{e}^{\circ} + RT\ln a_{ref}}
\]we derive:

\[\boxed{\mu_{e}(x,t) = \mu_{e}^{ref} + RT\ln\phi(x,t)}
\]Transposing, we get the interpretation for \(\phi:\)

\[\boxed{\phi(x,t) = \exp\left( \frac{\mu_{e}(x,t) - \mu_{e}^{ref}}{RT} \right)}\]

Further, we define observable electrical potential as:

\[\begin{matrix}
 & \boxed{V(x,t) = - \frac{RT}{F}\ln\phi(x,t)} & & 
\end{matrix}
\]

Where \(V(x,t)\): transmembrane potential (V, a two-point reading) and \(F\): Faraday constant (96485 C·mol\textsuperscript{-1})

\ul{Physical meaning of variable} \(\phi\): It quantifies the local capacity to hold or transfer electrons. High \(\phi\) indicates high electron availability / retention and low \(\phi\) means electron depletion. Since redox active species may be dynamically generated and consumed (electrons exist as bound states, radicals, redox intermediates), ϕ obeys a balance equation with source-sink terms, not a strict conservation law. It is an activity-based thermodynamic field reflecting electron availability and linked to chemical potential, that can also be used as a working variable for transport driven by \(\nabla\mu_{e}\) (as we shall see in sections to follow). Also, voltage is a derived observable arising from the underlying thermodynamic field \(\phi.\) Evolution follows reaction-transport, not pure continuity. A high EHP means the acquired electron is stabilized, residence time is long, and the species or site acts as an electron sink. On the contrary, a low EHP means the electron is labile, residence time is short, and the species/site may act as a relay or donor. The new terms of EDP (transient liberation or donating or e-generating), ERP (relaying or conducting or amplifying) and ESP (sinking or terminating or stabilizing) are the properties of species/sites. They replace terms like upstroke/downstroke and channel opening/closing type concepts, with ideas rooted in simple redox chemistry. The erstwhile HH model has no place for EHP, as it deals with ionic currents/capacitances/voltages and has no variable that represents electron residence time. EHP could (in effect!) be a function of the normalized concentration of ADRS, and is distinct from the standard redox potential \(E^{\circ}\), the electron affinity in vacuum and the band energy in solids. Instead, EHP is an emergent and contextual property that incorporates: intrinsic redox potential, solvation energy, dielectric environment, local electric fields, proximity of competing acceptors, spin signature, etc. We propose that EHP is the central state variable governing neuronal electrical behaviour. The present framework is built on the following physically and chemically grounded postulates:

1. Neuronal electrical phenomena are emergent properties of redox dynamics, not direct consequences of transmembrane ionic mass transfer.

2. Diffusible reactive species (DRS), including superoxide, hydroxide-associated anions, and redox-active semiquinones, are continuously generated (based on ECS) and quenched within the cellular milieu, and this process is absolutely essential for routine redox metabolism.

3. These DRS transiently store, relay, and dissipate electronic charge, giving rise to macroscopic ECS in the bulk aqueous-membrane environments.

4. The observed transmembrane potential (TMP) is a readout of asymmetry in EHP, rather than a driving variable.

5. Spatial signal transmission along axons occurs through electronic relay in the chemico-electromagnetic matrix (CEM), modulated by geometry, shielding, and redox kinetics.

These assumptions are collectively referred to as the murburn-thermodynamic framework, emphasizing mild, continuous redox activity in biological systems.

\subsection*{II. Local reaction-relaxation dynamics}

We first consider the evolution of the electron-holding field in the absence of spatial coupling (this is essential for developing the theory of spatial coupling later, and avoid double counting therein). Towards this, we define a local redox drive (target state):\(
\)\[\boxed{\phi_{\text{loc}}(\phi(x,t);x,t)}\]

as the instantaneous local steady-state value toward which the field \(\phi(x,t)\) relaxes under the prevailing biochemical conditions at position \(x\ \)and time \(t\). \(\phi_{\text{loc}}\) depends on local state and environment, and encodes: electron-donating processes (sources), electron-accepting processes (sinks) and catalytic redox cycling.

Then, we postulate that the time evolution of the electron holding field is governed by first-order relaxation dynamics (for sufficiently small relaxation time, τ):

\[\begin{matrix}
 & \boxed{\frac{\partial\phi(x,t)}{\partial t} = \frac{\phi_{\text{loc}}(\phi,x,t) - \phi(x,t)}{\tau(x)}} & & \text{(1)}
\end{matrix}
\]

Defining the auxiliary function \emph{f(ϕ):}

\[\boxed{f(\phi;x,t) \equiv \phi_{\text{loc}}(\phi,x,t) - \phi}
\]

So, the local dynamics become:

\[\boxed{\frac{\partial\phi}{\partial t} = \frac{1}{\tau(x)}\text{ }f(\phi;x,t)}
\]

The system relaxes toward a locally defined redox equilibrium. Now, we can interpret the relaxation time \(\tau(x)\) as:

\[\boxed{\tau(x)^{- 1} = \text{effective rate constant governing local redox equilibration}}
\]Determinants of \(\tau(x):\ \)reaction rate constants, enzyme concentrations, substrate availability, ionic strength, temperature, microenvironment, etc. Although the dynamics are local, parameters such as \(\tau(x)\) may vary with position due to enzyme distribution, cellular structure, and local biochemical environment.

Stability behavior: If \(\phi < \phi_{\text{loc}}\)→ \(\phi\ \)increases and if \(\phi > \phi_{\text{loc}}\)→ \(\phi\ \)decreases.

\(\phi_{\text{loc}}\)represents the net effect of local redox processes, and depends only on local state variables and contains no spatial derivatives.

Defining the steady (resting or fixed point) state \emph{ϕ}\textsubscript{rest}\hspace{0pt}

Fixed point (balanced condition of local steady state wherein production equals consumption):

\[\boxed{f(\phi_{\text{rest}};x,t) = 0} \Longleftrightarrow \phi_{\text{rest}} = \phi_{\text{loc}}(\phi_{\text{rest}},x,t)
\]\(\boxed{\phi_{\text{rest}}(x,t)\text{~is~the~fixed~point~of~the~local~dynamics}}\)which implies:

\[\boxed{\frac{\partial\phi}{\partial t} = 0}
\]

This state implies that no net redox imbalance exists and defines the resting or steady state at that location.

Define \(\mathbf{f}^{\mathbf{'}}\mathbf{(}\mathbf{\phi}_{\text{rest}}\mathbf{)}\)

\[\boxed{f^{'}(\phi) \equiv \frac{df}{d\phi} = \frac{d\phi_{\text{loc}}}{d\phi} - 1}
\]

Evaluated at the steady state:

\[\boxed{f^{'}(\phi_{\text{rest}}) = {\frac{d\phi_{\text{loc}}}{d\phi} \mid}_{\phi_{\text{rest}}} - 1}
\]

Linearization near the steady-state can be considered by a small perturbation:

\[\phi(x,t) = \phi_{\text{rest}} + \delta\phi(x,t), \mid \delta\phi \mid \ll 1\]

Expand \(f(\phi)\)about \(\phi_{\text{rest}}\)(Taylor series)

\[f(\phi_{\text{rest}} + \delta\phi) = f(\phi_{\text{rest}}) + f^{'}(\phi_{\text{rest}})\text{ }\delta\phi + \mathcal{O(}\delta\phi^{2})
\]

Then, at steady state, as f(\emph{ϕ}\textsubscript{rest}\emph{\hspace{0pt}}) = 0 (and approximating), we have:

\emph{f(ϕ) ≈ f′(ϕ\textsubscript{rest}\hspace{0pt}) δϕ}

Substitute into the evolution equation:

\[\frac{\partial(\phi_{\text{rest}} + \delta\phi)}{\partial t} = \frac{1}{\tau}\text{ }f^{'}(\phi_{\text{rest}})\text{ }\delta\phi
\]

\[\boxed{\frac{\partial(\delta\phi)}{\partial t} \approx \frac{1}{\tau}\text{ }f^{'}(\phi_{\text{rest}})\text{ }\delta\phi}
\]

where:

\[f(\phi) \equiv \phi_{\text{loc}}(\phi) - \phi
\]

\[{\boxed{f^{'}(\phi_{\text{rest}}) < 0 \Rightarrow \text{stable}}
}{\boxed{f^{'}(\phi_{\text{rest}}) > 0 \Rightarrow \text{unstable~(excitable)}}
}\]

In short, \(f(\phi)\ is\ \)the net local drive (source − sink), \(\phi_{\text{rest}}\)is the fixed point of local dynamics, \(f^{'}(\phi_{\text{rest}})\) determines stability/excitability and linearization is first-order Taylor expansion about the fixed point.

\ul{Justification of the linear relaxation form}: Equation (1) represents the lowest-order kinetic approximation consistent with: (i) Near-equilibrium chemical kinetics: rate \(\propto\) deviation from equilibrium, (ii) Linear response theory: flux or rate \(\propto\) thermodynamic driving force, and (iii) General relaxation processes: Analogous forms arise in dielectric relaxation, thermal equilibration, etc. The above equation is a first-order relaxation law, identical in structure to many well-known processes like RC circuit, chemical/thermal/spin relaxations, etc. Thus, the murburn equation is simply the redox analogue of other relaxation laws such as: chemical (concentration) relaxation: dC/dt = (C\textsubscript{eq} − C) / τ or thermal (temperature) relaxation: dT/dt = (T\textsubscript{env} − T) / τ

In connection to the reaction term in spatial formulation, we interpret Eqn. (1) as defining a local reaction rate \(R\ (x,\ t)\ or\ R_{loc}\), in a future section. It should be seen that the reaction term therein (within the spatial balance equation) is not independently postulated; it is derived from the local relaxation law. Local redox dynamics drive the system toward a state defined by \(\phi_{\text{loc}}\), with a characteristic time scale τ(x). We shall see that this local kinetic law provides the reaction term that, when combined with spatial transport, yields the full dynamical equation of neuronal electrical activity.

\subsection*{III. Spatial transport from thermodynamic gradients}

In Section II, the evolution of the electron-holding field was defined, without spatial coupling. Here, `transport' corresponds to propagation of redox/electronic (chemical potential) ripple, not physical displacement of mass/particles. We now generalize this description to a spatially distributed field,

\(\boxed{\phi(x,t):\ \text{electron-holding field defined over space and time}}\).

(i) Spatial dynamics arise because gradients in thermodynamic field \(\phi(x,t)\) drive transport. That is, spatial inhomogeneities in \emph{μ}\textsubscript{e} generate fluxes that act to equilibrate the redox field. From Section I: \(\mu_{e}(x,t) = \mu_{e}^{ref} + RT\ln\phi(x,t)\), we know \(\phi\ \)is a monotonic function of \(\mu_{e}\); therefore, gradients in \(\mu_{e}\)can be expressed via \(\phi.\ \)Taking gradient of the equation above, we have: ∇μ\textsubscript{e\hspace{0pt}} = \emph{RT}∇(ln\emph{ϕ}) \hspace{0pt} = \(RT\frac{\nabla\phi}{\phi}\) .

\[\boxed{\text{Spatial~inhomogeneities~in~}\mu_{e}\text{~generate~fluxes~that~act~to~equilibrate~the~redox~field}\text{.}}
\]

Transport occurs in response to gradients in chemical potential, driven by \(\nabla\mu_{e}(x,t)\). We define:

\[\begin{matrix}
 & \boxed{\mathbf{J}(x,t) = - M(x,t)\text{ }\nabla\mu_{e}(x,t)} & & \text{(2)}
\end{matrix}
\]\(where\ the\mathbf{\ }flux\ \mathbf{J}(x,t)\) is the net electron chemical potential transfer rate (or flux of redox/electronic influence) per unit area per unit time (a flux of redox/electronic influence, with units made self-consistent); and \(M(x,t)\) is a space-time dependent mobility coefficient (m\textsuperscript{2}·s\textsuperscript{-1})\(.\)

\(\ M\)depends on: redox carrier density, medium structure (water, lipid, proteins), ionic strength (electrostatic screening), temperature, geometry (axonal diameter, myelination), M ≈ constant under local homogeneity

\[\boxed{\text{Flux occurs from high chemical potential }\text{to}\text{ low chemical potential}}\]

Therefore:\(\begin{matrix}
 & \boxed{\mathbf{J}(x,t) = - M(x,t)\text{ }RT\text{ }\frac{\nabla\phi(x,t)}{\phi(x,t)}} & & \text{(3)}
\end{matrix}\)

To obtain a tractable transport law, we consider small deviations about a local reference value (\(\phi_{ref})\):

\[\phi(x,t) = \phi_{ref}(x,t) + \delta\phi(x,t), \mid \delta\phi \mid \ll \phi_{ref}\]

With the approximation: \(\frac{1}{\phi} \approx \frac{1}{\phi_{ref}}\) , define an effective transport coefficient:

\[\boxed{D(x,t) \equiv \frac{M(x,t)\text{ }RT}{\phi_{ref}(x,t)}}\]

\[\boxed{D(x,t)\text{ is an effective transport coefficient }\text{incorporating redox-mediated transfer mechanisms}\text{.}}
\]The unit of \emph{D} is m\textsuperscript{2}s\textsuperscript{-1} and depends on redox-active species density, molecular organization (water, proteins, lipids), ionic strength (screening effects), temperature, structural geometry (axon diameter, myelination). It represents electron/hole hopping, redox exchange, electrostatic coupling and medium-assisted relay. Substituting in Equation 3, the resulting flux law is:

\[\begin{matrix}
 & \boxed{\mathbf{J}(x,t) = - D(x,t)\text{ }\nabla\phi(x,t)} & & \text{(4)}
\end{matrix}
\]

The evolution of the field in space and time is then given by:

\[\begin{matrix}
 & \boxed{\frac{\partial\phi(x,t)}{\partial t} = - \nabla \cdot \mathbf{J}(x,t) + R(x,t)} & & \text{(5)}
\end{matrix}
\]Where \(\frac{\partial\phi}{\partial t}\): rate of change of the field, \(- \nabla \cdot \mathbf{J}\): net inflow/outflow due to transport, \(R(x,t)\): local reaction term. Equation 5 implies that transport redistributes the field and reactions create or annihilate electron-holding capacity.

Now, substituting Eqn. (4) into Eqn. (5), we have the evolution law:

\[\begin{matrix}
 & \boxed{\frac{\partial\phi}{\partial t} = \nabla \cdot \left( D(x,t)\text{ }\nabla\phi \right) + R(x,t)} & & \text{(6)}
\end{matrix}
\]Interpretation of spatial coupling: Spatial coupling arises solely through the divergence of flux ∇⋅(D∇ϕ). This is because local gradients \(\nabla\phi\) drive flux and spatial variation of flux (i.e., divergence) alters \(\phi\). Transport depends on gradients; local accumulation depends on curvature (via divergence). The above is identical in form to Fick's law, but \emph{D} is NOT simple diffusion, as it includes facilitated e-hopping, redox exchange and electrostatic mediation. The LHS is rate of change of thermodynamic field and the RHS has two terms: \(( - \nabla \cdot J)\), the net inflow of electron-equivalents and (\(R)\), the local creation/destruction via reactions. Although the structure of the equation appears to be similar to that of continuity-conservation, qualitative understanding (of the process involving continuous e-donation and e-sinking) tells us that it is a balance of flux plus local source/sink terms.

Special case: locally homogeneous medium: If \(D(x,t)\)varies slowly over space, then:

\[\boxed{\nabla \cdot (D\nabla\phi) \approx D\text{ }\nabla^{2}\phi}
\]Substituting the above in Eqn. 6:

\[\begin{matrix}
 & \boxed{\frac{\partial\phi}{\partial t} = D\text{ }\nabla^{2}\phi + R(x,t)} & & \text{(7)}
\end{matrix}\]

This approximation is valid over sufficiently small spatial scales or weak heterogeneity.

Spatial transport in the system is governed by gradients of a thermodynamically defined field, leading to fluxes that redistribute electron-holding capacity across space. The evolution of the field is determined by the balance between spatial transport and local reaction processes. Thus:

So, using the working variable (\(\phi = e^{\mu_{e}/RT}\)) and transport driver (\(\nabla\mu_{e}\)), we have arrived at the flux law (equation 4) and evolution profiles (equation 6).

\subsection*{IV. Unified reaction-diffusion-relaxation equation}

We introduced two complementary processes: `local reaction--relaxation dynamics' (Section II) and `spatial transport via flux divergence' (Section III). It can be registered that local dynamics create or remove \(\phi\) whereas transport redistributes it in space. The full dynamics is obtained by combining these.

From the mapping of Equation 1 of Section II and Equation 6 of Section III, we know that:

\[\begin{matrix}
 & \boxed{R(x,t) \equiv \frac{\phi_{loc}(\phi,x,t) - \phi(x,t)}{\tau(x)}} & & \text{(}\text{8}\text{)}
\end{matrix}
\]

This reaction term is required by the local relaxation law.

Substituting Eqn. (8) into equation 6:

\[\begin{matrix}
 & \boxed{\frac{\partial\phi(x,t)}{\partial t} = \nabla \cdot \left( D(x,t)\text{ }\nabla\phi(x,t) \right) + \frac{\phi_{loc}(\phi,x,t) - \phi(x,t)}{\tau(x)}} & & \text{(}\text{9}\text{)}
\end{matrix}
\]

The spatial coupling term \(\lbrack\nabla \cdot (D\nabla\phi)\rbrack\) represents relay of electronic ripple across space (spatial spread or relay) and depends on gradients and medium properties. The local reaction-relaxation term {[}\(\frac{\phi_{loc} - \phi}{\tau}\){]} governs local equilibration and encodes biochemical processes. The term on the left has the dimensions of T\textsuperscript{-1} (as \(\phi\) is dimensionless, \(\frac{\partial\phi}{\partial t}\) has the unit of s\textsuperscript{-1}) as do the two terms on the right. In the two limiting cases: (i) if there is no transport (\(D = 0\)) we have \(\text{pure relaxation}\text{,}\text{ }\text{∂ϕ}\text{\!/}\text{∂t }\text{= }\text{ }{(\phi}_{loc}\text{\!}\text{-}\text{ϕ}\text{)/}\tau\text{\!}\text{;}\) whereas (ii) if there is no reaction (\(\phi_{loc} = \phi;R = 0\)) we have \(\text{pure }\text{transport, }\text{∂ϕ}\text{\!/}\text{∂t }\text{=}\text{ }\text{∇}\text{⋅}\text{(}\text{D}\text{∇}\text{ϕ}\text{)}\). The evolution of the electron-holding field is governed by spatial transport driven by chemical potential gradients and local relaxation toward a redox-defined steady-state. Stated otherwise, the system evolves through the interplay of spatial redistribution of a thermodynamic field and local relaxation toward a redox-defined state. Equation 9 provides a complete description of the coupled spatial and local dynamics of the electron-holding field. All subsequent phenomena- waveform generation, propagation, threshold behavior, and conduction velocity- emerge from this unified equation. Recapitulating, \(D(x,t)\) controls propagation and spatial spread and depends on geometry, medium, ionic environment. \(\tau(x)\) sets temporal response scale. \(\phi_{loc}\) encodes local redox chemistry and will be further expanded in a later section. A specific reduced state of the equation valid over short spatial scales (or weak heterogeneity; using equation 7) is when \(D(x,t)\ \)and \(\tau(x)\) vary slowly over space. That is:

\[\boxed{D(x,t) \approx D,\tau(x) \approx \tau}
\]Then:

\[\begin{matrix}
 & \boxed{\frac{\partial\phi}{\partial t} = D\text{ }\nabla^{2}\phi + \frac{\phi_{loc} - \phi}{\tau}} & & \text{(10)}
\end{matrix}\]

It is reiterated that Equation 9 (\& 10) is a reaction-transport-relaxation relation governing a thermodynamic field rather than a conserved particle density.

\subsection*{V. Functional form of \(\phi_{\text{loc}}\) and the master murburn equation}

Equation 6 or 7 can be closed by defining \(\phi_{loc}\) as a function of \(\phi\), and this is required because \(\phi\) is the current local redox state and \(\phi_{loc}\) is the redox drive imposed by the same system locally.

The dependence of \(\phi\)\textsubscript{loc} on \(\phi\) could be linear or non-linear. Substitution as a purely linear form (such as \(\phi_{\text{loc}} = \Gamma_{0} + \lambda\phi\ \)into equation 9 would yield:

\[\frac{\partial\phi}{\partial t} = \nabla \cdot (D\nabla\phi) + \frac{\Gamma_{0} + (\lambda - 1)\phi}{\tau}.
\]

Purely linear dynamics and solutions only pose exponential growth or decay, with no saturation. Such systems cannot produce finite peaks (saturated spikes) or all-or-none (threshold) behavior and refractory periods. Therefore, non-linearity is an empirical requirement to justify the realities observed in neuronal function. To obtain bounded dynamics, growth term must be balanced by a non-linear saturation term. This is also justified because physically, finite redox capacity (saturation of electron-holding sites, recombination and quenching processes, competition between multiple redox pathways, etc.) is a reality. It can be envisaged that non-linearity can arise in a redox process by a higher (second) order function. We assume that:

\[\boxed{\phi_{\text{loc}}(\phi)\text{~is~a~smooth~scalar~function~of~the~scalar~variable~}\phi}
\]

We choose a reference \(\phi_{\text{ref}}\) (typically near the resting state):

\[\phi_{\text{loc}}(\phi) = \phi_{\text{loc}}(\phi_{\text{ref}}) + \phi_{\text{loc}}^{'}(\phi_{\text{ref}})\text{ }(\phi - \phi_{\text{ref}}) + \frac{1}{2}\phi_{\text{loc}}^{''}(\phi_{\text{ref}})\text{ }(\phi - \phi_{\text{ref}})^{2} + \cdots
\]

This is just the standard Taylor series expansion. Now, we define a shifted variable:

\[\boxed{u \equiv \phi - \phi_{\text{ref}}}
\]Now the expansion becomes:

\[\phi_{\text{loc}} = A + B\text{ }u + C\text{ }u^{2} + \cdots
\]

where: \(A = \phi_{\text{loc}}(\phi_{\text{ref}})\), \(B = \phi_{\text{loc}}^{'}(\phi_{\text{ref}})\), \(C = \frac{1}{2}\phi_{\text{loc}}^{''}(\phi_{\text{ref}})\)

In Section III: \(\phi_{\text{ref}}\ \)used to linearize \(\nabla\mu_{e} \sim (RT/\phi_{\text{ref}})\nabla\phi\) and in Section V, it was used as expansion point for \(\phi_{\text{loc}}(\phi).\ \)Both instances are valid if \(\phi_{\text{ref}}\  \approx \text{local~operating~(resting)~state}\ \) \(value\ about\ which\ perturbations\ are\ small.\)

We define new notations (for convenient handling):

\[\boxed{\Gamma_{0} \equiv A,\lambda \equiv B,\beta \equiv - C}
\]

\begin{quote}
\(\Gamma_{0}\): basal redox drive (constant offset)

\(\lambda\): linear gain or amplification (excitability)

\(\beta > 0\): nonlinear saturation coefficient
\end{quote}

Because we want physical saturation, we opt for the negative sign for the quadratic term:

\[\boxed{\phi_{\text{loc}}^{''}(\phi_{\text{ref}}) < 0\text{\:\,} \Rightarrow \text{\:\,}\beta > 0}
\]

\[\boxed{\mid u \mid = \mid \phi - \phi_{\text{ref}} \mid \text{~is~not~too~large}}
\]

Then: linear term signifies growth; quadratic term gives saturation and higher-order terms are smaller corrections. Therefore, quadratic order is the minimal nonlinearity that produces bounded dynamics. This is exactly the same logic used in Landau expansions, reaction--diffusion systems and nonlinear kinetics.

Then:

\[\boxed{\phi_{\text{loc}} = \Gamma_{0} + \lambda u - \beta u^{2}}\]

Now, we can understand the emergence of excitability in the system by substituting \(\phi_{loc} = \Gamma_{0} + \lambda\phi - \beta\phi^{2}\) into Equations 9. we get:

\[\boxed{\frac{\partial\phi}{\partial t} = \nabla \cdot (D\nabla\phi) + \frac{\Gamma_{0} + (\lambda - 1)\phi - \beta\phi^{2}}{\tau}}\ \ \ \ \ \ \ \ \text{(11)}\]

For the special case of expanding equation 7 (the locally homogeneous form), we get:

\[\frac{\partial\phi}{\partial t} = D\nabla^{2}\phi + \frac{\Gamma_{0} + (\lambda - 1)\phi - \beta\phi^{2}}{\tau}\]

\ul{Interpretation of parameters}

\(\lambda\): excitability (controls linear amplification, determines instability threshold)

\(\beta\): saturation (limits growth, stabilizes system)

\(\Gamma_{0}\): basal drive (sets baseline activity, linked to metabolic/redox conditions)

\(D\): transport coefficient (governs spatial spread)

\(\tau\): time scale (controls response speed)

\ul{Stability and excitability}

Linear stability (small perturbations): \(\frac{\partial\phi}{\partial t} \approx D\nabla^{2}\phi + \frac{(\lambda - 1)\phi}{\tau}\)

Instability condition: \(\boxed{\lambda > 1 \Rightarrow \text{growth of perturbations}}\)

Saturation or stabilization: \(- \beta\phi^{2}\text{limits growth}\)

Steady state (resting value): Setting \(\frac{\partial\phi}{\partial t} = 0,\nabla^{2}\phi = 0\)

\begin{quote}
Solving for: \(\Gamma_{0} + (\lambda - 1)\phi - \beta\phi^{2} = 0\)

Gives: \(\boxed{\phi_{rest} = \frac{(\lambda - 1) + \sqrt{\left( \lambda - 1)^{2} + 4\beta\Gamma_{0} \right.\ }}{2\beta}}\)

Corresponding voltage: \(V_{rest} = - \frac{RT}{F}\ln\phi_{rest}\)
\end{quote}

Nonlinearity in local redox dynamics generates excitability, threshold behavior, and saturation within a single unified framework. 

The resulting reaction-diffusion-relaxation equation constitutes the master equation (written in its expanded form below) governing neuronal electrical dynamics. As a single scalar reaction--diffusion field it supports travelling fronts and relaxation to rest; the self-resetting pulse and refractory period of a full action potential additionally require a recovery variable, which we introduce in Section~XI.

\[\boxed{\frac{\partial\phi}{\partial t} = \nabla \cdot (D\nabla\phi) + \frac{\Gamma_{0} + (\lambda - 1)\phi - \beta\phi^{2}}{\tau}}
\]

\subsection*{VI. Non-dimensionalization of the murburn master equation}

To reduce the number of independent parameters and identify the fundamental dimensionless groups governing system behavior, we need to non-dimensionalize the murburn master equation 11. Non-dimensionalization helps in identifying the minimal parameter set controlling the system dynamics.First, we choose characteristic scales, as below:

(i) Time scale: \(\boxed{t = \tau_{0}\text{ }t^{'}}\)

\(\tau_{0}\): characteristic relaxation time (reference value of \(\tau(x)\))

\(t^{'}\): dimensionless time

(ii) Length scale: \(\boxed{x = L\text{ }x^{'}}\)

\(L\): characteristic spatial scale (e.g., axonal length scale)

\(x^{'}\): dimensionless position

(iii) Field variable: \(\boxed{\phi = \phi_{ref}\text{ }u(x^{'},t^{'})}\)

\(\phi_{ref}\): characteristic magnitude (e.g., resting value scale)

\(u\): dimensionless field variable (when \(u = 1\), we have the reference/resting state; avoids conflict with the earlier instance of Taylor expansion where we deemed \(u \equiv \phi - \phi_{\text{ref}}\))

Then, transformation of derivatives would be:

(i) Time derivative: \(\frac{\partial\phi}{\partial t} = \frac{\phi_{ref}}{\tau_{0}}\frac{\partial u}{\partial t^{'}}\)

(ii) Spatial derivatives: \(\nabla\phi = \frac{\phi_{ref}}{L}\nabla^{'}u\) ; \(\nabla^{2}\phi = \frac{\phi_{ref}}{L^{2}}\nabla^{'2}u\)

Further, we substitute this into master equation 11:

\(\frac{\phi_{ref}}{\tau_{0}}\frac{\partial u}{\partial t^{'}} =\) \(\frac{\phi_{\text{ref}}}{L^{2}}\nabla^{'} \cdot (D(Lx^{'})\nabla^{'}u) + \frac{\Gamma_{0} + (\lambda - 1)\phi_{ref}u - \beta\phi_{ref}^{2}u^{2}}{\tau}\)

Simplifying and dividing through by \(\phi_{ref}/\tau_{0},\) we have:

\(\frac{\partial u}{\partial t^{'}} = \nabla^{'} \cdot \left( \frac{D\tau_{0}}{L^{2}}\text{ }\nabla^{'}u \right) + \frac{\tau_{0}}{\tau}\left\lbrack \frac{\Gamma_{0}}{\phi_{ref}} + (\lambda - 1)u - \beta\phi_{ref}u^{2} \right\rbrack\)

\ul{Let's define the dimensionless parameters}

Transport parameter: \(\boxed{D(x^{'}) \equiv \frac{D(Lx^{'})\text{ }\tau_{0}}{L^{2}}}\); \(\boxed{x = Lx^{'}\  \Rightarrow \ D(x) \rightarrow D(Lx^{'})}\)

Time-scale ratio: \(\boxed{\Theta(x^{'}) \equiv \frac{\tau_{0}}{\tau(x)}}\)

Basal drive: \(\boxed{\gamma \equiv \frac{\Gamma_{0}}{\phi_{ref}}}\)

Linear feedback parameter: \(\boxed{\alpha \equiv (\lambda - 1)}\)

Nonlinearity parameter: \(\boxed{\beta^{'} \equiv \beta\text{ }\phi_{ref}}\)

Then, the dimensionless master equation becomes:

\[\begin{matrix}
 & \boxed{\frac{\partial u}{\partial t^{'}} = \nabla^{'} \cdot (D(x^{'})\text{ }\nabla^{'}u) + \Theta(x^{'})\left\lbrack \gamma + \alpha u - \beta^{'}u^{2} \right\rbrack} & & \text{(12)}
\end{matrix}
\]

\ul{Interpretation of dimensionless groups}

\(\mathcal{D}\): transport strength (controls spatial spread, influences propagation velocity)

\(\Theta\): responsiveness (ratio of global to local time scales, strength of reaction vs transport)

\(\alpha\): excitability parameter (\(\alpha > 0\): amplifying regime, \(\alpha < 0\): damping regime)

\(\beta^{'}\): saturation strength (limits growth, stabilizes waveform)

\(\gamma\): basal drive (sets baseline state, reflects metabolic/redox input)

Special case of simplified homogeneous form (parameters are \textasciitilde constant):

If parameters are approximately constant:

\[\begin{matrix}
 & \boxed{\frac{\partial u}{\partial t^{'}} = \mathcal{D}\text{ }\nabla^{'2}u + \Theta\left( \gamma + \alpha u - \beta^{'}u^{2} \right)} & & \text{(13)}
\end{matrix}
\]

Conceptual significance: The parameter space is governed by the independent dimensionless parameters of: \(\mathcal{D}\)\emph{, Θ, α, β\textquotesingle, γ}. As a consequence, different physiological systems map onto the same parameter space, which enables comparison across conditions. Non-dimensionalization reveals that neuronal dynamics are governed by the balance between transport, responsiveness, amplification, saturation and basal drive. This reduced form provides the basis for analyzing waveform shape, propagation velocity, and parameter sensitivity in subsequent sections.

\subsection*{VII. Resting state and range of trans-membrane potential}

At rest, the system is approximately stationary and spatially uniform (with little transport), with negligible temporal variation and spatial gradients.

\(\boxed{\phi(x,t) \approx \phi_{\text{rest}} = \text{constant}}\)The resting state corresponds to a stationary and spatially uniform condition, defined by:

\[\boxed{\frac{\partial u}{\partial t^{'}} = 0,\nabla^{'2}u = 0}
\]The above follows from spatial uniformity. The resting state is determined solely by local reaction-relaxation balance.

From the dimensionless equation 13, and from \(\boxed{\frac{\partial u}{\partial t^{'}} = 0,\nabla^{'2}u = 0},\ \)we have:

\[\begin{matrix}
 & \boxed{\gamma + \alpha u - \beta^{'}u^{2} = 0} & & \text{(1}\text{4}\text{)}
\end{matrix}
\]The physically admissible solution corresponds to the positive root ensuring \emph{u}\textsubscript{rest} \textgreater{} 0

Solution for resting state is: \(\boxed{u_{rest} = \frac{\alpha\  + \ \sqrt{\alpha^{2} + 4\beta^{'}\gamma}}{2\beta^{'}}}\)

We apply physical constraint \(\boxed{u_{rest} > 0}\) (consistent with activity-based definition of \(\phi\)) and convert to dimensional field,

\[\phi = \phi_{ref}\text{ }u
\]Resting value:

\[\begin{matrix}
 & \boxed{\phi_{rest} = \phi_{ref}\text{ }\frac{\alpha\  + \ \sqrt{\alpha^{2} + 4\beta^{'}\gamma}}{2\beta^{'}}} & & \text{(1}\text{5}\text{)}
\end{matrix}
\]

Resting transmembrane potential: From Section I (by definition of the electrochemical mapping):

\[V = - \frac{RT}{F}\ln\phi
\]Resting voltage

\[\begin{array}{c}
\boxed{V_{rest} = - \frac{RT}{F}\ln\left\lbrack \phi_{ref}\text{ }\frac{\alpha + \sqrt{\alpha^{2} + 4\beta^{'}\gamma}}{2\beta^{'}} \right\rbrack} \qquad \text{(16)} \\[2.5ex]
\boxed{\text{Existence~of~a~real~positive~solution~requires~}\alpha^{2} + 4\beta^{'}\gamma > 0}
\end{array}\]

\ul{Physiological interpretation:} V\textsubscript{rest} is determined by the interplay of basal drive, feedback strength and saturation, as:

\(\gamma\): increases baseline redox drive → raises \(\phi_{rest}\)→ more negative \(V\)

\(\alpha\): controls linear amplification (excitability)

\(\beta^{'}\): limits buildup via saturation

\ul{Range of resting potentials}: \(\boxed{V_{rest} \approx - 10\text{ mV to } - 100\text{ mV}}\)

Table 1 presents a display of some corresponding \(\phi\)values using: \(\phi = e^{- V/(RT/F)} \approx e^{- V/25.7}\)

Table 1: Relation of resting voltage with the field variable \(\mathbf{\phi}\)

\begin{longtable}[]{@{}
  >{\raggedright\arraybackslash}p{(\columnwidth - 2\tabcolsep) * \real{0.5821}}
  >{\raggedright\arraybackslash}p{(\columnwidth - 2\tabcolsep) * \real{0.4179}}@{}}
\toprule\noalign{}
\begin{minipage}[b]{\linewidth}\raggedright
\(\mathbf{V}_{\mathbf{rest}}\)(mV)
\end{minipage} & \begin{minipage}[b]{\linewidth}\raggedright
\[\mathbf{\phi}_{\mathbf{rest}}\]
\end{minipage} \\
\midrule\noalign{}
\endhead
\bottomrule\noalign{}
\endlastfoot
-10 & \textasciitilde1.48 \\
-30 & \textasciitilde3.21 \\
-50 & \textasciitilde6.997 \\
-70 & \textasciitilde15.23 \\
-90 & \textasciitilde33.18 \\
\end{longtable}

The resting transmembrane potential emerges as a stable fixed point of nonlinear redox dynamics. Although ions influence the redox states in cells, it is seen that they are not the primary rationales for observed voltage. The resting membrane potential emerges as a stable fixed point of the nonlinear redox dynamics. Small fluctuations around \emph{ϕ}\textsubscript{rest} correspond to baseline neuronal activity. This framework links biochemical parameters directly to measurable voltage without invoking ion-specific transport laws as primary determinants.

\subsection*{VIII. Excitability, all-or-none response, spike initiation, and basic signal propagation}

Local reactions govern evolution and excitability. From the dimensionless equation 12 or 13, we know the local reaction term to be:

\[\boxed{f(u) = \gamma + \alpha u - \beta^{'}u^{2}}\ \ \ \ \ \ \ \ \ \ \ \ (14a)\]

It determines fixed points and stability and is independent of spatial transport and the resting state satisfies: \(f(u_{rest}) = 0\). So,

\[f^{'}(u) = \frac{df}{du} = \alpha - 2\beta^{'}u
\]Stability requires: \(\boxed{\alpha - 2\beta^{'}u_{rest} < 0}\); Nonlinearity ensures a bounded dynamics and stable resting state is determined by \(f^{'}\left( u_{\text{rest}} \right) < 0(as\ shown\ below)\)!

(i) Let's now explore the threshold for excitation and consider small perturbations.

Starting from the dimensionless equation (local + transport):

\[\frac{\partial u}{\partial t^{'}} = \nabla^{'} \cdot (D\nabla^{'}u) + \Theta f(u)
\]

Step 1: perturb about the steady state

\[u = u_{\text{rest}} + \delta u, \mid \delta u \mid \ll 1
\]

Step 2: expand \(f(u)\ \)(Taylor expansion)

\[f(u_{\text{rest}} + \delta u) = f(u_{\text{rest}}) + f^{'}(u_{\text{rest}})\text{ }\delta u + \mathcal{O(}\delta u^{2})
\]

But:

\[f(u_{\text{rest}}) = 0
\]

So:

\[f(u) \approx f^{'}(u_{\text{rest}})\text{ }\delta u
\]

Step 3: substitute into PDE

\[\frac{\partial(u_{\text{rest}} + \delta u)}{\partial t^{'}} = \nabla^{'} \cdot (D\nabla^{'}\delta u) + \Theta f^{'}(u_{\text{rest}})\delta u
\]

Since \(u_{\text{rest}}\)is constant:

\[\frac{\partial\delta u}{\partial t^{'}} = \nabla^{'} \cdot (D\nabla^{'}\delta u) + \Theta f^{'}(u_{\text{rest}})\delta u
\]

For local threshold analysis, we ignore spatial variation:

\[\boxed{\nabla^{'} \cdot (D\nabla^{'}\delta u) \approx 0}
\]

Therefore, linearized dynamics:

\[{\boxed{\frac{\partial(\delta u)}{\partial t^{'}} \approx \Theta\text{ }f^{'}(u_{\text{rest}})\text{ }\delta u}
}
\]Now, we have two regimes:

(a) Subthreshold perturbation\(;\) \(\boxed{f^{'}(u_{\text{rest}}) < 0\text{\:\,} \Rightarrow \text{\:\,}\delta u \rightarrow 0};\) System returns to rest.

(b) Suprathreshold perturbation; \(\boxed{f(u) > 0}\); then \(\delta u\text{ grows}\text{, and perturbation increases}\).

\[\begin{array}{c}
\boxed{\text{Threshold~is~the~smallest~}u_{\text{th}}\text{~such~that~}f(u_{\text{th}}) = 0\text{~and~}f^{'}(u_{\text{th}}) > 0} \\[2.5ex]
\boxed{\begin{array}{c}
\text{The~resting~state~is~stable~if~}f^{'}(u_{\text{rest}}) < 0. \\
\text{A~perturbation~grows~only~if~it~crosses~into~a~region~where~}f(u) > 0.
\end{array}}
\end{array}\]Linear stability conveys what happens near rest; threshold is determined by the nonlinear condition \emph{f (u)} = 0.

(ii) The mechanism for all-or-none response

Once \(u\ \)crosses threshold: linear term (\(\alpha u\)) amplifies and nonlinear term (\(- \beta^{'}u^{2}\)) limits growth. The system therefore, evolves to a finite amplitude excursion, independent of initial perturbation size. Response amplitude is determined by system parameters, not by input magnitude. With this quadratic (monostable) reaction the local dynamics simply relax back to rest, with no threshold and no all-or-none response; a genuine threshold requires a bistable (cubic) reaction, and a self-resetting spike with a refractory period requires in addition a recovery variable, which Section~XI introduces.

(iii) Spike initiation

Initiation by a localized perturbation: \(u(x^{'},t^{'}) = u_{rest} + \Delta u(x^{'})\)

Condition for spike initiation: \(\boxed{\exists\text{ }x^{'}\text{ such that }f(u) > 0}\)

Evolution: local amplification occurs, saturation limits peak, recovery follows due to negative feedback; all of it resulting in a transient localized excursion (spike) emerges from nonlinear reaction dynamics.

Role of spatial transport (for convenience, taking equation 13): \(\frac{\partial u}{\partial t^{'}}\mathcal{= D}\text{ }\nabla^{'2}u + \Theta f(u)\)\\
Mechanism of propagation: Reaction term creates spike locally and the diffusion term spreads perturbation. That is: spatial transport couples neighboring regions, enabling relay of excitation. A spike propagates if local amplification exceeds diffusive dissipation. That is: if D is too small, excitation remains localized and does not spread; but if amplification is sufficient, the signal travels. The overall dynamics is governed by reaction--transport balance, as minimally represented by:

\[\text{Propagation} \sim \frac{\text{growth rate}}{\text{diffusion rate}}
\]This balance determines whether a traveling wave is sustained.

(iv) Traveling wave concept (qualitative) can be minimally conceived in the form:

\(u(x^{'},t^{'}) = U(z),z = x^{'} - v^{'}t^{'}\)wherein the waveform moves without changing shape and \(v^{'}\): dimensionless propagation velocity. Then the governing equation would be:

\(\mathcal{D}\text{ }\frac{d^{2}U}{dz^{2}} + v^{'}\frac{dU}{dz} + \Theta f(U) = 0\) (17)\(
\)(since: \(\frac{\partial z}{\partial x'} = 1,\ \ \ \ \ \ \frac{\partial z}{\partial t'} = \  - v^{'}\ \ \ \  \Rightarrow \ \ \ \ \ \frac{\partial u}{\partial t^{'}} = - v'\frac{\partial U}{\partial z}\),\(\ \ \ \ \ \frac{d^{2}u}{d{x'}^{2}} = \ \frac{d^{2}U}{dz^{2}}\))

\[\boxed{\text{This~is~a~second-order~nonlinear~ODE~governing~wave~shape~and~velocity}\text{.}}
\]

The synthesis exercise till now has demonstrated that- (i) excitability arises from nonlinear local dynamics, while propagation arises from spatial coupling, (ii) all-or-none response, spike initiation, and signal propagation emerge from a single reaction-transport framework; no separate mechanisms are required for thresholding and conduction.

\emph{D} too small

\[\boxed{\text{No~spatial~coupling~→~spike~stays~localized~→~no~propagation}}
\]

\emph{D} too large

\[\boxed{\text{Signal~spreads~too~fast~→~amplitude~diluted~→~spike~dies}}
\]

Intermediate \emph{D} (correct regime)

\[\boxed{\text{Spread~is~sufficient,~but~amplification~dominates~→~traveling~wave}}
\]Therefore:

\[\boxed{\begin{array}{r}
\text{Excitability~arises~from~nonlinear~local~reaction~dynamics,~while} \\
\text{propagation~arises~from~spatial~coupling~of~the~thermodynamic~field.} \\
\text{All-or-none~response,~spike~initiation,~and~signal~propagation~emerge} \\
\text{within~a~unified~reaction–transport~framework.}
\end{array}}
\]

\subsection*{IX. Quantitative regimes for baseline activity, graded responses, and action potentials}

From equations 13 and 14, we can gauge that all dynamical behaviors arise from the interplay of \(f(u)\) and spatial coupling. That is:\\
\[\frac{\partial u}{\partial t^{'}} = \nabla^{'} \cdot (D\nabla^{'}u) + \Theta f(u),f(u) = \gamma + \alpha u - \beta^{'}u^{2}
\]Now, we move on to specific domains of neuronal activity, as measured experimentally.

(i) Baseline activity (subthreshold dynamics): It corresponds to fluctuations near the stable fixed point: \(u \approx u_{rest}\) (linearized dynamics)

Let: \(u = u_{rest} + \delta u,\  \mid \delta u \mid \ll 1;Then\ linearized\ dynamics\ (earlier\ section)is:
\)\[\frac{\partial(\delta u)}{\partial t^{'}}\mathcal{\approx D}\text{ }\nabla^{'2}\delta u + \Theta f^{'}(u_{rest})\text{ }\delta u
\]If \(f^{'}(u_{rest}) < 0\): perturbations decay; if \(f^{'}\left( u_{rest} \right) \approx 0\): slow fluctuations persist.

Baseline firing arises from weak fluctuations around a stable but near-marginal state. Small stochastic or environmental perturbations could have incomplete damping and this may produce low-level activity.

(ii) Graded responses/potentials: It occurs when perturbations remain below excitation threshold; \(f(u) < 0\ \text{for all }u\text{ reached}\). Response amplitude may be proportional to input but no regenerative amplification occurs and spatial spread governed by diffusion.

Approximate equation: \(\frac{\partial u}{\partial t^{'}}\mathcal{\approx D}\text{ }\nabla^{'2}u + \Theta(\gamma + \alpha u)\)

Amplitude depends continuously on input magnitude and signal attenuates with distance due to transport without amplification. Local redox perturbation is insufficient dominated by linear response and passive spread.

(iii) Transition to action potential: It happens when threshold is crossed.

When a perturbation satisfies: \(\boxed{f(u) > 0\ \text{over a finite region}}
\)The local amplification begins and the nonlinear term becomes active; and the system transitions from graded/passive to regenerative dynamics.

(iv) Generation of the action potential waveform

Ignoring spatial coupling at a point: \(\frac{du}{dt^{'}} = \Theta\left( \gamma + \alpha u - \beta^{'}u^{2} \right)
\)\ul{Phase-wise dynamics}

(a) Resting phase: \(u \approx u_{rest},f(u) = 0
\)(b) Depolarization (rising phase): \(f(u) > 0\); linear term dominates, rapid increase in \(u\)

(c) Peak saturation: \emph{f} (\emph{u}) \emph{=} 0\emph{~}at~peak;\hspace{0pt} \(- \beta^{'}u^{2}\text{ balances growth}\); peak amplitude is determined by the balancing of α and β\textquotesingle{}\(
\)(d) Repolarization: \(f(u) < 0\text{ for large }u\); nonlinear term dominates, decay begins

(e) Recovery: \(uu_{rest}\)

The full spike waveform emerges from intrinsic nonlinear reaction dynamics. Fast rise connotes a linear amplification and slower decay indicates a nonlinear saturation.

\ul{Quantitative features of waveform}

Amplitude: \(\boxed{u_{\text{max}} \sim \frac{\alpha}{\beta^{'}}}
\)Rise time: \(\boxed{t_{\text{rise}} \sim \frac{1}{\Theta\text{ }\alpha}}\) or \(\boxed{t_{\text{rise}} \sim \frac{1}{\Theta\text{ } \mid f^{'}(u_{\text{rest}}) \mid}}\)\\
Decay time: \(\boxed{t_{\text{decay}} \sim \frac{1}{\Theta\text{ }\beta^{'}u}}\ or\ \boxed{t_{\text{decay}} \sim \frac{1}{\Theta\text{ }\beta^{'}u_{\text{max}}}}\)

\[
{\boxed{\text{Fast~rise~and~slower~decay~arise~naturally~from~asymmetric~nonlinear~dynamics}\text{.}}
}\]

\ul{Spatial structure of action potential}

Full equation: \(\boxed{\frac{\partial u}{\partial t^{'}} = \nabla^{'} \cdot (D\nabla^{'}u) + \Theta f(u)}\)

Local reaction generates spike and diffusion/transport spreads excitation, with neighboring regions cross threshold. The spike propagates as a self-sustained traveling disturbance. Table 2 gives a demarcation of the various regimes for neuronal electricality.

Table 2: Distinction between various regimes

\begin{longtable}[]{@{}
  >{\raggedright\arraybackslash}p{(\columnwidth - 4\tabcolsep) * \real{0.3171}}
  >{\raggedright\arraybackslash}p{(\columnwidth - 4\tabcolsep) * \real{0.2964}}
  >{\raggedright\arraybackslash}p{(\columnwidth - 4\tabcolsep) * \real{0.3865}}@{}}
\toprule\noalign{}
\begin{minipage}[b]{\linewidth}\raggedright
Regime
\end{minipage} & \begin{minipage}[b]{\linewidth}\raggedright
Condition
\end{minipage} & \begin{minipage}[b]{\linewidth}\raggedright
Behavior
\end{minipage} \\
\midrule\noalign{}
\endhead
\bottomrule\noalign{}
\endlastfoot
Baseline & near \(u_{rest}\) & small fluctuations \\
Graded & \(f(u) < 0\) & proportional response \\
Action potential & \(f(u) > 0\)locally & regenerative spike \\
\end{longtable}

Once again, all three regimes above emerge from the same governing equation (dynamical regimes of the same nonlinear reaction-diffusion/transport-relaxation system without additional mechanisms. Transitions between these regimes are governed by the sign and magnitude of the local reaction term \emph{f}(\emph{u}). Waveform shape and dynamics arise intrinsically from the balance between amplification, saturation, and transport.

\subsection*{X. Neuronal conduction velocity (NCV): formula and magnitude}

In this framework/model, the signal is not a voltage pulse per se, but a traveling disturbance in the field \emph{u}(\emph{x′,t′}) (equivalently ϕ). So, the problem is- if a spike forms at one location, how fast does that disturbance move to neighboring regions? Instead of solving the full PDE everywhere, we approach the solution based on physically motivated observations: spikes maintain a characteristic shape and they move through space at a roughly constant speed. This brings us to the traveling wave ansatz.

We seek solutions of the form: \(\boxed{u(x^{'},t^{'}) = U(z),z = x^{'} - v^{'}t^{'}}\)

The above means that the waveform moves with constant shape (simply shifts in space) and \(v^{'}\) is the dimensionless propagation velocity.

\[\boxed{U\text{~}\text{is an unbounded but dynamically limited field; not a probability or normalized variable.}}\]

\[\boxed{U = 1 \Rightarrow \phi = \phi_{\text{ref}}\ (\text{reference/resting~scale})}
\]

\[\boxed{U_{\max}\text{~is~not~fixed;~determined~by~}f(U) = 0\ ( \sim \alpha/\beta^{'})}
\]

Transformation of derivatives (converting the PDE to ODE):

\[{\frac{\partial u}{\partial t^{'}} = - v^{'}\frac{dU}{dz}
}{\nabla^{'2}u = \frac{d^{2}U}{dz^{2}}
}\]Substitution into governing equation: \(- v^{'}\frac{dU}{dz} = \mathcal{D}\frac{d^{2}U}{dz^{2}} + \Theta\left( \gamma + \alpha U - \beta^{'}U^{2} \right)\)

Rearranged form gives: \(\begin{matrix}
 & \boxed{\mathcal{D}\frac{d^{2}U}{dz^{2}} + v^{'}\frac{dU}{dz} + \Theta\left( \gamma + \alpha U - \beta^{'}U^{2} \right) = 0} & & \text{(}\text{17a}\text{)}
\end{matrix}\)

Now instead of tracking space and time separately, we analyze how the shape \emph{U(z)} behaves as it moves. The entire velocity selection happens at the leading edge of the wave. At the front, the signal is just beginning to rise; amplitudes are very small and nonlinear saturation has not yet kicked in. So, the system behaves approximately linearly there.

Linearization at the leading edge (or front): At the wave front, the system is near resting state and only a small perturbation exists. That perturbation either: grows leading to wave propagation OR decays, which results in the wave dying out. So, the speed of propagation is controlled by how fast small perturbations grow and spread.

Let

\[U(z) = u_{\text{rest}} + \delta U(z), \mid \delta U \mid \ll 1
\]

Linearization about \(\mathbf{u}_{\text{rest}}\)

Using \(f\left( u_{\text{rest}} \right) = 0\ \)and Taylor expansion (like earlier!):

\[f(U) \approx f^{'}(u_{\text{rest}})\text{ }\delta U
\]

Substituting into equation 17a above:

\[\begin{matrix}
 & \boxed{D\frac{d^{2}\delta U}{dz^{2}} + v^{'}\frac{d\delta U}{dz} + \Theta f^{'}(u_{\text{rest}})\text{ }\delta U = 0} & & \text{(1}\text{8}\text{)}
\end{matrix}
\]The resting state is stable, so \(f^{'}(u_{\text{rest}}) < 0\).

We define a positive growth rate ahead of the front: \(\boxed{\sigma \equiv - f^{'}(u_{\text{rest}}) > 0}\)

Then Eqn. (18) becomes: \(\begin{matrix}
 & \boxed{D\frac{d^{2}\delta U}{dz^{2}} + v^{'}\frac{d\delta U}{dz} - \Theta\sigma\text{ }\delta U = 0} & & \text{(18a)}
\end{matrix}\)

Exponential ansatz

Assume a decaying front: \(\boxed{\delta U(z) = e^{- \kappa z},\kappa > 0}\)

This means that the front decays smoothly ahead of the spike; no oscillations or discontinuities. This is the standard physical behavior of a stable propagating front.

Substitution: \(\mathcal{D}\kappa^{2}e^{- \kappa z} - v^{'}\kappa e^{- \kappa z} + \Theta f^{'}(u_{rest})e^{- \kappa z} = 0\)

Cancelling exponential, we get:

\[\begin{matrix}
 & \boxed{D\kappa^{2} - v^{'}\kappa - \Theta\sigma = 0} & & \text{(19)}
\end{matrix}
\]For a physically admissible solution: \(\kappa > 0\ \text{and real}\).

Marginal (slowest stable) front selection fixes the speed through \mbox{\(v^{'2} \geq 4\mathcal{D}\Theta\sigma\)} (with \mbox{\(\sigma \equiv -f^{'}(u_{rest}) > 0\)}), the equality giving the selected value; this marginal-stability condition presumes an excitable (linearly unstable) leading edge.

Minimum velocity condition: \(\begin{matrix}
 & \boxed{{v'}_{\min}^{} = 2\sqrt{D\text{ }\Theta\text{ }\sigma}} & & \text{(}\text{20}\text{)}
\end{matrix}\); it is the selected propagation velocity.

The minimum velocity is taken because: slower waves cannot sustain themselves (perturbation decays), faster waves are mathematically possible but unstable and the system naturally selects the minimum stable velocity. This is the Fisher--KPP velocity-selection idea; it applies cleanly when the invaded state is linearly unstable, so for the stable resting axon the expression is read as a marginal/scaling estimate.

Expression for growth rate: \(f^{'}(u) = \alpha - 2\beta^{'}u\)

At resting state: \(\boxed{f^{'}(u_{rest}) = \alpha - 2\beta^{'}u_{rest}}\); therefore, \(\boxed{\sigma = 2\beta^{'}u_{\text{rest}} - \alpha}\)

Using steady-state condition: \(\gamma + \alpha u_{rest} - \beta^{'}u_{rest}^{2} = 0\)

\[\boxed{f^{'}(u_{rest})\text{ captures local excitability near rest}}
\]Conversion to dimensional velocity from dimensionless NCV:

\[\begin{matrix}
 & \boxed{{v'}_{\min}^{} = 2\sqrt{\mathcal{D}\Theta\sigma}} & & 
\end{matrix}\]

Recall: \(x = Lx^{'},t = \tau_{0}t^{'}\); Therefore: \(\boxed{v = \frac{L}{\tau_{0}}v^{'}}\)

So, \(\boxed{v = \frac{2L}{\tau_{0}}\sqrt{\mathcal{D}\Theta\sigma}}\)

Substitute definitions: \(\mathcal{D =}\frac{D\tau_{0}}{L^{2}},\Theta = \frac{\tau_{0}}{\tau}\)

The simplified form is: \(\begin{matrix}
 & \boxed{v' = 2\sqrt{\frac{D}{\tau}\text{ }\sigma}} & & \text{(}\text{20a}\text{)}
\end{matrix}\)\\
\ul{Interpretation of NCV formula}

\(\boxed{v' \propto \sqrt{D}}\); faster transport or faster propagation; how fast wave front spreads, includes: redox relay, medium structure, shielding

\(\boxed{v' \propto \frac{1}{\sqrt{\tau}}}\); faster local kinetics or faster conduction (smaller \(\tau,\ \)faster spread); how fast local system responds

\(\boxed{v' \propto \sqrt{f^{'}\left( u_{rest} \right)}\ or\ \sqrt{\sigma}}\ \)stronger excitability or faster signal (bigger value, faster spread); how strongly the system amplifies the small perturbations

\[\boxed{\parbox{0.86\linewidth}{\centering Signal velocity is governed by the balance between transport, kinetics, and local growth near threshold.}}\]

\ul{Physiological magnitude}

\emph{D} in this framework represents an effective electron-relay transport coefficient, which can exceed molecular diffusion by several orders of magnitude due to field-assisted hopping and network-mediated transfer processes.

Typical values: \(D \sim 10^{- 7}\text{–}10^{- 3}\text{ }\text{m}^{2}/\text{s}\); \(\tau \sim 10^{- 4}\text{–}10^{- 2}\text{ s}\); \(\sigma \sim 0.1\text{–}10\)

If we take: \(D \sim 10^{- 5}\)to \(10^{- 3}\text{ }\text{m}^{2}/\text{s}\); \(\tau \sim 10^{- 4}\); \(f^{'} \sim 1\text{–}10\)

Estimated NCV: \(\boxed{1\ to\ 100\text{ m/s}}\); which spans the range of observed neuronal conduction velocities; this is an order-of-magnitude consistency over a wide parameter window, not yet a parameter-free prediction.

\subsection*{XI. Excitable extension: recovery variable, refractoriness, and a selected pulse speed}

Sections VIII--X established that the single scalar field \(u\) yields a stable resting state and graded responses, while its quadratic (monostable) reaction supports travelling fronts and relaxation but not a genuine threshold, a self-resetting all-or-none spike, or a refractory period; nor could a sharp conduction speed be selected from a stable rest state (Section X). Following the FitzHugh--Nagumo route, we retain the murburn redox interpretation and restore full excitability by coupling the EHP field \(u\) to a slow recovery variable \(w\):
\[
\frac{\partial u}{\partial t^{'}} = \mathcal{D}\,\nabla^{'2}u + \Theta\left\lbrack R(u) - w \right\rbrack,\qquad
\frac{\partial w}{\partial t^{'}} = \varepsilon\,(u - \delta\,w),\qquad \varepsilon \ll 1 .
\]
Here \(u\) is the fast EHP field of the preceding sections, \(w\) is a slow recovery field, \(\varepsilon\) sets the time-scale separation (slow recovery), and \(\delta>0\) fixes the recovery decay. The local reaction \(R(u)\) is taken in its bistable, N-shaped (cubic) form: the quadratic \(\gamma + \alpha u - \beta^{'}u^{2}\) of Sections II and V is the leading expansion of the local redox kinetics, and retaining the next, saturating order supplies the second stable branch that a true threshold requires.

Physically, \(w\) represents the slow recovery of the local one-electron (DROS/oxygen--EHP) pool. A spike transiently depletes the reactive pool; the \(-\Theta w\) term then opposes the drive, producing repolarisation and a refractory interval during which re-excitation is suppressed until the pool replenishes on the slow scale \(\varepsilon^{-1}\). This is the murburn counterpart of the slow recovery (\(K^{+}\)/gating relaxation) that the recovery variable encodes in FitzHugh--Nagumo and Hodgkin--Huxley.

With the recovery variable the model is genuinely excitable: (i) a finite threshold separates sub- from supra-threshold perturbations; (ii) a supra-threshold input triggers a fixed-amplitude all-or-none excursion in \(u\) before \(w\) returns the system to rest; (iii) the elevated \(w\) sets the absolute and relative refractory period (\(\sim \varepsilon^{-1}\)); and (iv) the spatial system now admits a travelling \emph{pulse} (not merely a front), invading the stable rest state, whose speed is the discrete eigenvalue of the travelling-pulse problem. The scaling \(v^{'}\sim 2\sqrt{(\mathcal{D}/\tau)\,\sigma}\) of Eq.~(20a) then fixes the order-of-magnitude pulse speed, while the recovery variable selects a unique stable pulse and removes the front-selection ambiguity of Section X. In the limit \(\varepsilon\to 0\) with \(w\) held fixed, the fast subsystem reduces to a single-field reaction--diffusion equation of the Sections VIII--X type, with the quadratic \(\gamma+\alpha u-\beta^{'}u^{2}\) recovered as the near-rest expansion of \(R(u)\); the resting-state, graded-response and velocity-scaling results of those sections then carry over as the fast-time limit.

\subsection*{XII. Physiological and experimental determinants of waveform and NCV}

Having derived the master equation and conduction velocity

\[\boxed{v' = 2\sqrt{\frac{D}{\tau}\text{ }\sigma},\sigma \equiv - f^{'}(u_{\text{rest}}) > 0}
\]

\[\boxed{\text{Both~waveform~and~propagation~arise~from~the~same~parameter~set:~}D,\ \tau,\ \alpha,\ \beta^{'},\ \gamma}
\]

we now examine how biological factors influence neuronal behavior through their impact on model parameters of: \(\mathcal{D}\)\emph{,} \(\tau\)\emph{, α, β\textquotesingle,} and \emph{γ} (Θ is not really an independent parameter, as it is contingent upon \(\tau)\). (i) Axonal diameter: larger diameter leads to reduced resistance to transport, increased effective coupling; axonal diameter primarily increases D. (ii) Myelination: reduces transverse (radial) dissipation, enhances longitudinal coupling, increases effective relay efficiency; Myelination increases effective D and reduces loss of signal. Minimal amplitude change. (iii) Ionic strength: modulates \(D\) through electrostatic screening, affects interaction between redox species. Effect on transport: moderate screening facilitates relay, while excessive screening suppresses coupling Effect on waveform: alters spatial spread, influences spike width (iv) Temperature dependence: increases reaction rates, enhances mobility, reduces viscosity. Parameter influence: Temperature decreases τ and increases D; as a consequence, higher temperature leads to faster NCV and sharper waveforms. (v) Redox environment and metabolic state Basal drive \(\gamma\): reflects metabolic input, determines resting state. Effects: increased metabolism results in higher \(\gamma\), shifts \(u_{rest}\)

Consequence: Changes in γ alter resting potential and excitability

(vi) Effect of parameters

\[\begin{array}{c}
\boxed{D\text{ controls “how far \& how quick” the signal spreads, not “how tall” it is}} \\[1ex]
\boxed{\tau \uparrow \Rightarrow \text{slower dynamics → broader spikes}} \\[1ex]
\boxed{\alpha \uparrow \Rightarrow \text{faster rise + higher peak}} \\[1ex]
\boxed{\beta^{'} \uparrow \Rightarrow \text{lower amplitude + sharper cutoff}} \\[1ex]
\boxed{\gamma \uparrow \Rightarrow \text{higher baseline → reduced dynamic range}}
\end{array}\]

\(\ \)(vii) Impact on waveform characteristics

Amplitude

\[\boxed{u_{\max} \sim \frac{\alpha}{\beta^{'}}}
\]Rise time

\[\boxed{t_{\text{rise}} \sim \frac{1}{\Theta\text{ } \mid f^{'}(u_{\text{rest}}) \mid} = \frac{1}{\Theta\text{ }\sigma}}
\]Decay / recovery

\[\boxed{t_{\text{decay}} \sim \frac{1}{\Theta\text{ }\beta^{'}u_{\max}}}
\]

Spike width

\[\boxed{t_{\text{width}} \sim t_{\text{rise}} + t_{\text{decay}}}
\]

\[\boxed{\text{Waveform~asymmetry~(fast~rise,~slower~decay)~emerges~intrinsically}}
\]

In short, waveform morphology and conduction velocity are governed by distinct but coupled parameter groups. Transport parameter (\(D\)) leads to spatial propagation, Kinetic parameters (\(\tau,\alpha,\beta^{'}\)) determine waveform shape, and basal drive (\(\gamma\)) determines the operating point. Table 3 presents a predictive table for the impact of various factors (environmental parameters or intrinsic features) in the murburn model.

Table 3: Theoretical and intuitive predictions/indications of the murburn model of neuronal electricality

\begin{longtable}[]{@{}
  >{\raggedright\arraybackslash}p{(\columnwidth - 12\tabcolsep) * \real{0.1709}}
  >{\raggedright\arraybackslash}p{(\columnwidth - 12\tabcolsep) * \real{0.1691}}
  >{\raggedright\arraybackslash}p{(\columnwidth - 12\tabcolsep) * \real{0.0721}}
  >{\raggedright\arraybackslash}p{(\columnwidth - 12\tabcolsep) * \real{0.1251}}
  >{\raggedright\arraybackslash}p{(\columnwidth - 12\tabcolsep) * \real{0.0984}}
  >{\raggedright\arraybackslash}p{(\columnwidth - 12\tabcolsep) * \real{0.1257}}
  >{\raggedright\arraybackslash}p{(\columnwidth - 12\tabcolsep) * \real{0.2387}}@{}}
\toprule\noalign{}
\begin{minipage}[b]{\linewidth}\raggedright
Factor
\end{minipage} & \begin{minipage}[b]{\linewidth}\raggedright
Parameter affected
\end{minipage} & \begin{minipage}[b]{\linewidth}\raggedright
NCV
\end{minipage} & \begin{minipage}[b]{\linewidth}\raggedright
Amplitude
\end{minipage} & \begin{minipage}[b]{\linewidth}\raggedright
Rise Time
\end{minipage} & \begin{minipage}[b]{\linewidth}\raggedright
Spike Width
\end{minipage} & \begin{minipage}[b]{\linewidth}\raggedright
Shape Effect
\end{minipage} \\
\midrule\noalign{}
\endhead
\bottomrule\noalign{}
\endlastfoot
Axonal diameter & D↑ & ↑ & \textasciitilde{} & \textasciitilde{} & ↓ (slightly) & sharper propagation front \\
Myelination & D↑, losses ↓ & ↑↑ & \textasciitilde{} & ↓ & ↓ & narrower, cleaner spikes \\
Temperature ↑ & D↑, τ↓ & ↑ & slight ↑ & ↓↓ & ↓↓ & faster, sharper spikes \\
Ionic strength ↑ & D (nonlinear) & ± & \textasciitilde{} & ± & ± & can broaden or damp \\
Metabolic drive ↑ & γ↑ → urest↑ & ± & slight ↓ & slight ↑ & ↑ & broader baseline, reduced contrast \\
Excitability ↑ & α↑ & ↑ & ↑↑ & ↓ & ↓ & steeper, higher spikes \\
Saturation ↑ & β′↑ & ↓ & ↓ & \textasciitilde{} & ↓ & clipped, smaller spikes \\
Reaction slowdown & τ↑ & ↓ & \textasciitilde{} & ↑ & ↑↑ & sluggish, broader spikes \\
\end{longtable}

Both conduction velocity and waveform characteristics emerge from a single unified parameter set, providing a direct bridge between physiological conditions and measurable neuronal behavior, wherein transport parameters (D) primarily determine propagation velocity; kinetic parameters (τ,α,β′) determine waveform morphology; and basal drive (γ) sets the operating point and dynamic range.

\[\boxed{\begin{array}{r}
\text{The~neuronal~signal~is~a~propagating~redox-field~disturbance~rather~than} \\
\text{a~purely~ionic~current~pulse.}
\end{array}}
\]

Neuronal conduction velocity emerges from the interplay of spatial transport and local nonlinear amplification. Velocity is not an independent parameter but is determined by intrinsic system properties. The same equation governing waveform also determines propagation speed.

The derivation does not add artificial gating variables or permeability coefficients, and the parameters/variables used have tangible/viable physical contexts. We only assume natural and feasible (and well-documented) factual postulates:

\begin{quote}
1. ECS-DRS-CEM redox chemistry generates charge.

2. Charge relaxes toward chemical equilibrium while it spreads locally in ms.
\end{quote}

From those two minimal assumptions, the ϕ--τ equation emerges. We do not treat the neuronal axon as a uniformly laid out pipe of ions with well-distributed membrane pumps and channels, but consider it as a redox field. At rest, chemistry holds electrons in balance. When redox drive shifts briefly, electron-holding capacity collapses locally. That produces a voltage spike and because electron imbalance spreads along the axon, the spike propagates. All of this follows from one reaction-relaxation--diffusion equation. Excitability emerges when redox feedback exceeds redox relaxation. The neuron is excitable and responds to all types of stimuli because it is thermodynamically, kinetically and mechanistically sensitive to oxygen-superoxide redox equilibrium.

\section*{4. Summation}

The reaction-diffusion framework underlying murburn dynamics suggests that excitability and signal propagation can be interpreted as emergent properties of redox-mediated charge generation, relay, and dissipation. In biological systems, membranes and proteins provide essential control over selectivity, kinetics, spatial confinement, and energetic coupling, thereby enabling the precision, robustness, and adaptability observed in living neurons. Redox active proteins like neuroglobin could be especially suited to serve in the electron-buffering and stochastic relay process within this framework.

EHP (ϕ) is a dimensionless field variable related logarithmically to electron chemical potential. It is the effective electron-holding capacity that correlates to the population of redox-active species. Its evolution therefore follows a reaction-relaxation-transport equation, and not a purely conservative continuity law. Perturbations in redox balance, including shifts in oxygen-superoxide equilibrium, can modulate this field and thereby influence neuronal responses. Stimuli give CEM perturbation and O₂/O₂•⁻ equilibrium shifts. This affects the EHP (EDP-ERP-ESP) network, which in turn culminates in a suitable neuronal response. The murburn theory assigns chemically grounded causal roles and directionality, enforces thermodynamic consistency, and allows cross-system generalization (mitochondria, neurons, photoreceptors). It also affords seamless integration from metabolism to TMP and perturbation, to neuronal conduction, and to cognition. We postulate that the oxygen-superoxide equilibrium thus serves as the aerobic~cellular antenna, translating all forms of environmental energy into the common language of redox potential; a language that neurons have evolved to interpret with exquisite sensitivity and specificity. The basic concepts and events in the murburn model are captured in Figure 1. The unified master equation derived herein accounts for resting potential, threshold behavior, spike waveform, and signal propagation within a single minimal formalism. It predicts that conduction velocity and waveform morphology are governed by transport efficiency, local reaction kinetics, and nonlinear feedback, thereby enabling direct mapping from physiological variables to measurable electrical behavior. This perspective complements existing electrophysiology models and provides a coherent bridge from molecular redox chemistry to cellular electrophysiology. It also offers a predictive framework for understanding neuronal dynamics, which shall be compared and elaborated in another part of this communication.

\includegraphics[width=6.26806in,height=3.52569in]{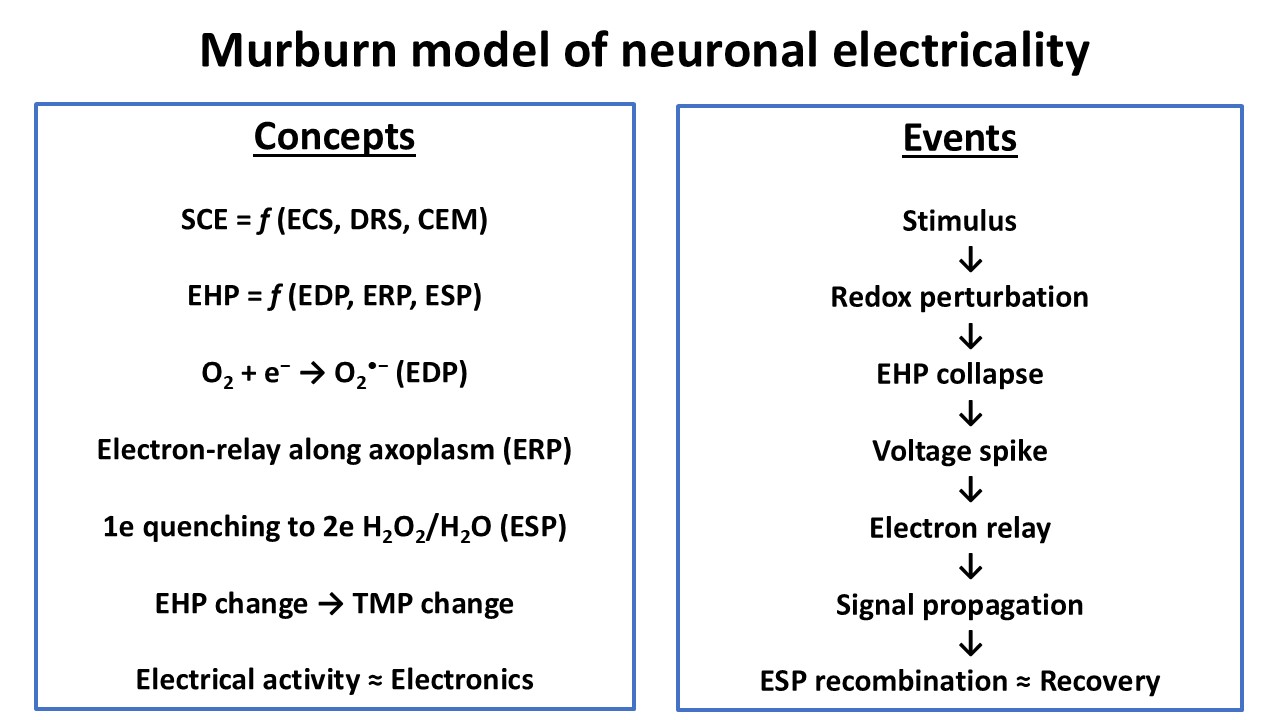}

\emph{\textbf{Figure 1: Schema of terms and concepts in the murburn model neuronal electrical activity.}}

\textbf{Disclaimers:} KMM wrote the first draft of the paper and theorized the overall layout. NS found errors and lacunae, and corrected text. TH validated the murburn model with simulations, provided constructive inputs and minor editorial corrections. MK provided critical inputs. AA carried out the mathematical development and refinement of the model and prepared the revised manuscript. This work was powered by Satyamjayatu: The Science \& Ethics Foundation.

\section*{References}
\begin{refslist}
Armstrong, C. M. (1971) Interaction of tetraethylammonium ion derivatives with the potassium channels of giant axons. \emph{The Journal of General Physiology}, 58(4), 413-437. doi: 10.1085/jgp.58.4.413

Cahalan MD, (1978) Local anesthetic block of sodium channels in normal and pronase-treated squid giant axons. \emph{Biophysical Journal}. 23(2):285-311. doi: 10.1016/S0006-3495(78)85449-6.

Carter BC, Bean BP. (2009) Sodium entry during action potentials of mammalian neurons: incomplete inactivation and reduced metabolic efficiency in fast-spiking neurons. Neuron. 2009 Dec 24;64(6):898-909. doi: 10.1016/j.neuron.2009.12.011.

DeMaegd ML, Städele C, Stein W. (2017) Axonal Conduction Velocity Measurement. \emph{Bio Protoc.} 2017 Mar 5;7(5): e2152. doi: 10.21769/BioProtoc.2152.

El Hady, A., \& Machta, B. B. (2015). Mechanical surface waves accompany action potential propagation. \emph{Nature Communications}, 6, 6697.

Fatt, P. \& Katz, B. (1953). The electrical properties of crustacean muscle fibres. \emph{Journal of Physiology}, 120(1-2), 171--204. doi: 10.1113/jphysiol. 1953.sp004884.

FitzHugh, R. (1961). Impulses and physiological states in theoretical models of nerve membrane. \emph{Biophysical Journal}, 1, 445--466.

Gideon DA, Nirusimhan V, Edward JC, Sudarsha K, Manoj KM. (2022) Mechanism of electron transfers mediated by cytochromes c and b5 in mitochondria and endoplasmic reticulum: classical and murburn perspectives. \emph{Journal of Biomolecular Structure and Dynamics}. 40 (19), 9235-9252. doi: 10.1080/07391102.2021.1925154.

Heimburg, T., \& Jackson, A. D. (2005) On soliton propagation in biomembranes and nerves. \emph{Proceedings of the National Academy of the Sciences USA}, 102, 9790--9795.

Hodgkin, A. L., \& Huxley, A. F. (1952). A quantitative description of membrane current and its application to conduction and excitation in nerve.~\emph{The Journal of Physiology}, 117(4), 500-544.~

Jaeken L, Manoj KM. (2025) Murburn Bioenergetics and ``Origins--Sustenance--Termination--Evolution of Life'': Emergence of Intelligence from a Network of Molecules, Unbound Ions, Radicals and Radiations. \emph{International Journal of Molecular Sciences} 26 (15), 7542.

Kandel, E. R., et al. (2021).~\emph{Principles of Neural Science}~(6th ed.). McGraw Hill.

Llinás, R., \& Sugimori, M. (1980). Electrophysiological properties of Guillain-Barré Purkinje cells in vitro. \emph{Journal of Physiology}, 305, 171-195. ~doi: 10.1113/jphysiol. 1980.sp013358\\
Mangoni ME, Nargeot J. (2008) Genesis and regulation of the heart automaticity. Physiological Reviews. 88(3):919-82. doi: 10.1152/physrev.00018.2007.

Manoj KM, Bazhin N, Tamagawa H. (2022b) The murburn precepts for cellular ionic homeostasis and electrophysiology. \emph{Journal of Cellular Physiology}. 237(1):804-814. doi: 10.1002/jcp.30547.

Manoj KM, Bazhin NM. (2021) The murburn precepts for aerobic respiration and redox homeostasis. \emph{Progress in Biophysics and Molecular Biology.} 167:104-120. doi: 10.1016/j.pbiomolbio.2021.05.010.

Manoj KM, et al. (2023b) Murburn concept in cellular function and bioenergetics, Part 1: Understanding murzymes at the molecular level. \emph{AIP Advances}, 13, 120702.

Manoj KM, et al. (2023c) Murburn concept in cellular function and bioenergetics, Part 2: Understanding integrations-translations from molecular to macroscopic levels. \emph{AIP Advances}, 13, 120701.

Manoj KM, et al. (2026b) Algorithmic identification of murzymes and murburn mechanisms based on structural, theoretical, experimental and generic features. \emph{BioMed Research International}. DOI: 10.1155/bmri/2577941

Manoj KM, et al. (2026c) Structure and mechanism of cellular cation-transporters: Affinity-binding and murburn models. \emph{International Journal of Biological Macromolecules}. DOI: 10.1016/j.ijbiomac.2026.150614

Manoj KM, Gideon DA, Bazhin NM, Tamagawa H, Nirusimhan V, Kavdia M, Jaeken L. (2023a) Na,K‐ATPase: A murzyme facilitating thermodynamic equilibriums at the membrane‐interface\emph{. Journal of Cellular Physiology} 238 (1), 109-136. doi: 10.1002/jcp.30925

Manoj KM, Gideon DA, Samuel PM, Das S. (2026a) Quantitative Treatments for Explaining the Mechanism and Kinetics of Catalytic Electron Transfers in Murburn Processes, Particularly Involving Heme Enzymes Like (Per) oxidases and P450s. \emph{BioMed Research International} (1), 3079294.

Manoj KM, Jaeken L (2023) Synthesis of theories on cellular powering, coherence, homeostasis and electro-mechanics: Murburn concept \& evolutionary perspectives. \emph{Journal of Cellular Physiology}, 238, 931-953. doi: 10.1002/jcp.31000

Manoj KM, Nirusimhan V, Parashar A, Edward J, Gideon DA. (2022a) Murburn precepts for lactic-acidosis, Cori cycle, and Warburg effect: Interactive dynamics of dehydrogenases, protons, and oxygen. \emph{Journal of Cellular Physiology}. 237(3):1902-1922. doi: 10.1002/jcp.30661.

Manoj KM, Parashar A, Gade SK, Venkatachalam A. (2016) Functioning of Microsomal Cytochrome P450s: Murburn Concept Explains the Metabolism of Xenobiotics in Hepatocytes. \emph{Frontiers in Pharmacology} 7:161. doi: 10.3389/fphar.2016.00161.

Manoj KM, Tamagawa H, Bazhin N, Jaeken L, Nirusimhan V, Faraci F, Gideon DA. (2022c) Murburn model of vision: Precepts and proof of concept. \emph{Journal of Cellular Physiology}. 237(8):3338-3355. doi: 10.1002/jcp.30786.

Manoj KM, Tamagawa H. (2022) Critical analysis of explanations for cellular homeostasis and electrophysiology from murburn perspective. \emph{Journal of Cellular Physiology}. 237(1):421-435. doi: 10.1002/jcp.30578.

Neher, E., \& Sakmann, B. (1976). Single-channel currents recorded from membrane of denervated frog muscle fibres.~\emph{Nature}, 260, 799--802.

Parashar A, Jacob VD, Gideon DA, Manoj KM. (2022) Hemoglobin catalyzes ATP-synthesis in human erythrocytes: a murburn model. \emph{Journal of Biomolecular Structure and Dynamics}. 40 (19), 8783-8795. doi: 10.1080/07391102.2021.1925592.

Vandael, D. H., et al. (2013). CaV1.3 channels in adrenal chromaffin cells and their role in the pacemaker potential. Cell Calcium, 54(2), 105--115. doi: 10.1016/j.ceca.2013.04.006
\end{refslist}

\end{document}